\documentclass{elsarticle}

\usepackage[reviewcopy]{adndt}
\usepackage{longtable}

\usepackage{amsmath}

\usepackage{amssymb}

\biboptions{square,sort&compress}
\bibpunct[]{[}{]}{,}{n}{}{;}
\begin{document}

\begin{frontmatter}

\journal{Atomic Data and Nuclear Data Tables}


\title{Discovery of Cesium, Lanthanum, Praseodymium and Promethium Isotopes}

\author{E. May}
\author{M. Thoennessen\corref{cor1}}\ead{thoennessen@nscl.msu.edu}

 \cortext[cor1]{Corresponding author.}

 \address{National Superconducting Cyclotron Laboratory and \\ Department of Physics and Astronomy, Michigan State University, \\ East Lansing, MI 48824, USA}

\begin{abstract}
Currently, forty-one cesium, thirty-five lanthanum, thirty-two praseodymium, and thirty-one promethium, isotopes have been observed and the discovery of these isotopes is discussed here. For each isotope a brief synopsis of the first refereed publication, including the production and identification method, is presented.
\end{abstract}

\end{frontmatter}





\newpage
\tableofcontents
\listofDtables

\vskip5pc

\section{Introduction}\label{s:intro}

The discovery of cesium, lanthanum, praseodymium and promethium isotopes is discussed as part of the series summarizing the discovery of isotopes, beginning with the cerium isotopes in 2009 \cite{2009Gin01}. Guidelines for assigning credit for discovery are (1) clear identification, either through decay-curves and relationships to other known isotopes, particle or $\gamma$-ray spectra, or unique mass and Z-identification, and (2) publication of the discovery in a refereed journal. The authors and year of the first publication, the laboratory where the isotopes were produced as well as the production and identification methods are discussed. When appropriate, references to conference proceedings, internal reports, and theses are included. When a discovery includes a half-life measurement the measured value is compared to the currently adopted value taken from the NUBASE evaluation \cite{2003Aud01} which is based on the ENSDF database \cite{2008ENS01}. In cases where the reported half-life differed significantly from the adopted half-life (up to approximately a factor of two), we searched the subsequent literature for indications that the measurement was erroneous. If that was not the case we credited the authors with the discovery in spite of the inaccurate half-life. All reported half-lives inconsistent with the presently adopted half-life for the ground state were compared to isomers half-lives and accepted as discoveries if appropriate following the criterium described above. If the first observation of an isotope corresponded to an isomeric state, the first observation of the ground state is also included.

The first criterium excludes measurements of half-lives of a given element without mass identification. This affects mostly isotopes first observed in fission where decay curves of chemically separated elements were measured without the capability to determine their mass. Also the four-parameter measurements (see, for example, Ref. \cite{1970Joh01}) were, in general, not considered because the mass identification was only $\pm$1 mass unit.

The second criterium affects especially the isotopes studied within the Manhattan Project. Although an overview of the results was published in 1946 \cite{1946TPP01}, most of the papers were only published in the Plutonium Project Records of the Manhattan Project Technical Series, Vol. 9A, ``Radiochemistry and the Fission Products,'' in three books by Wiley in 1951 \cite{1951Cor01}. We considered this first unclassified publication to be equivalent to a refereed paper.

The initial literature search was performed using the databases ENSDF \cite{2008ENS01} and NSR \cite{2008NSR01} of the National Nuclear Data Center at Brookhaven National Laboratory. These databases are complete and reliable back to the early 1960's. For earlier references, several editions of the Table of Isotopes were used \cite{1940Liv01,1944Sea01,1948Sea01,1953Hol02,1958Str01,1967Led01}. A good reference for the discovery of the stable isotopes was the second edition of Aston's book ``Mass Spectra and Isotopes'' \cite{1942Ast01}.

\section{Discovery of $^{112-152}$Cs}

Forty-one cesium isotopes from A = 112--152 have been discovered so far; these include 1 stable, 21 neutron-deficient and 19 neutron-rich isotopes. According to the HFB-14 model \cite{2007Gor01,2007HFB01}, $^{180}$Cs should be the last odd-odd particle stable neutron-rich nucleus while the odd-even particle stable neutron-rich nuclei should continue through $^{185}$Cs. The proton dripline has already been reached and two isotopes beyond the dripline have been identified as proton emitters. It is anticipated that four more isotopes ($^{108-111}$Cs) could still have half-lives longer than 10$^{-9}$~ns \cite{2004Tho01}. Thus, about 35 isotopes have yet to be discovered corresponding to 46\% of all possible cesium isotopes.

\begin{figure}
	\centering
	\includegraphics[scale=.7]{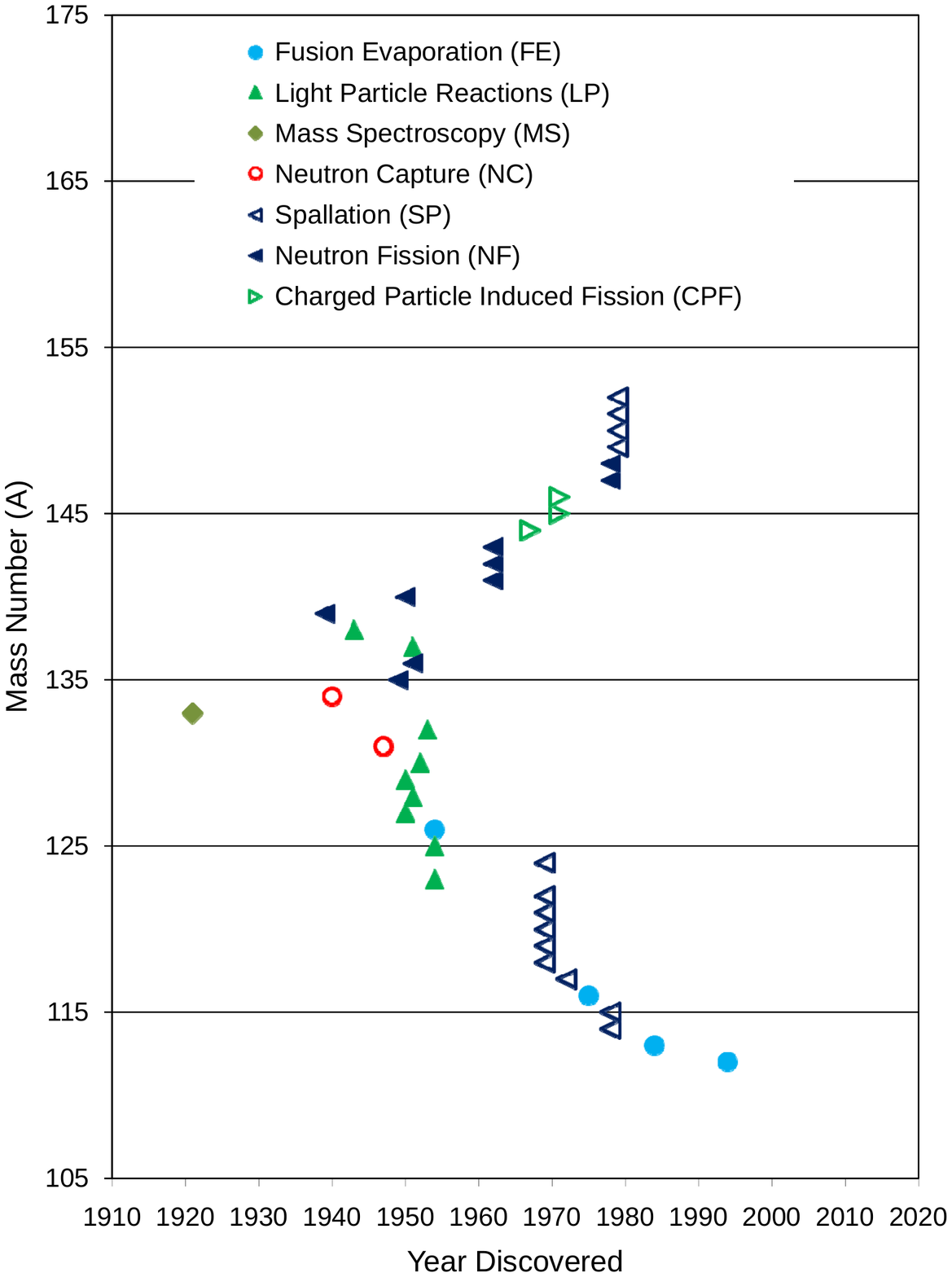}
	\caption{Cesium isotopes as a function of time when they were discovered. The different production methods are indicated.}
\label{f:year-cs}
\end{figure}

Figure \ref{f:year-cs} summarizes the year of first discovery for all cesium isotopes identified by the method of discovery. The range of isotopes predicted to exist is indicated on the right side of the figure. The radioactive cesium isotopes were produced using fusion evaporation reactions (FE), light-particle reactions (LP), neutron capture (NC) neutron induced fission (NF), charged-particle induced fission (CPF), and spallation (SP). The stable isotope was identified using mass spectroscopy (MS). The discovery of each cesium isotope is discussed in detail and a summary is presented in Table 1.

\subsection*{$^{112}$Cs}\vspace{0.0cm}
``Decays of odd-odd \emph{N - Z} = 2 nuclei above $^{100}$Sn: The observation of proton radioactivity from $^{112}$Cs'' was published in 1994 reporting the discovery of $^{112}$Cs by Page et al.\cite{1994Pag01}. A 259 MeV $^{58}$Ni beam bombarded an isotopically enriched $^{58}$Ni target and $^{112}$Cs was produced via the \emph{p3n} evaporation channel. The isotopes were separated with the Daresbury Recoil Mass Separator and implanted in a double-sided silicon strip detector. ``In accordance with expected alpha decay branching ratios, a correlation analysis identified three events in this group consistent with the decay chain: $^{112}$Cs $\stackrel{p}{\longrightarrow}$ $^{111}$Xe $\stackrel{\alpha}{\longrightarrow}$ $^{107}$Te $\stackrel{\alpha}{\longrightarrow}$, establishing beyond doubt that this \emph{A=}112 activity represents the first observation of the proton decay of the new isotope $^{112}$Cs.'' The measured half-life of 500(100)~$\mu$s corresponds to the currently accepted value. A previous search for the proton decay of $^{112}$Cs was unsuccessful \cite{1991Hei02}.

\subsection*{$^{113}$Cs}\vspace{0.0cm}
Faestermann et al. published ``Evidence for proton radioactivity of $^{113}$Cs and $^{109}$I'' in 1984 describing the first observation of $^{113}$Cs \cite{1984Fae01}. Enriched $^{58}$Ni targets were bombarded with a 250 Mev $^{58}$Ni beam from the Munich MP Tandem-linear accelerator combination. Evaporation residues were collected on a catcher foil and charged particles were measured with a parallel plate avalanche counter and a Bragg curve spectroscopy ionization chamber. ``With this in mind we can conclusively assign the 0.98~MeV proton radioactivity to $^{113}$Cs populated with the $^{58}$Ni($^{58}$Ni,p2n) reaction.'' The reported half-life of 0.9$^{+1.3}_{-0.4}~\mu$s was later corrected to 33(7)~$\mu$s \cite{1987Gil01}: ``The half-life of $^{113}$Cs of the order of 1 $\mu$s,which was observed in our first measurement, could not be confirmed by measurements with improved statistics. The limited statistics of the first measurement however had left a finite probability that the half-life was long compared to the 1.4~$\mu$s long interval subtended by the time spectrum.'' The later value is barely within a factor of two of the presently accepted value of 16.7(7)$\mu$s. The proton energy was measured correctly and these data were accepted as the first observation of $^{113}$Cs in the subsequent literature \cite{1994Pag01,1998Bat01}. Earlier searches for $^{113}$Cs were unsuccessful \cite{1978DAu01}.

\subsection*{$^{114,115}$Cs}\vspace{0.0cm}
$^{114}$Cs and $^{115}$Cs were discovered in 1978 by D'Auria et al. in the 1978 paper ``Properties of the lightest known cesium isotopes $^{114-118}$Cs'' \cite{1978DAu01}. A lanthanum target was bombarded with 600 MeV protons from the CERN synchrocyclotron. $^{114}$Cs and $^{115}$Cs were formed in spallation reactions and identified with ISOLDE isotope separator.  ``Since the daughter, $^{114}$Xe, has been shown not to be a delayed particle emitter, the observed activity is attributed to $^{114}$Cs... The delayed-proton activity exhibited a half-life of about 1 sec for $^{115}$Cs, but the much stronger proton branch of the daughter made a precise determination difficult. An experiment with set-up A gave a more accurate value: 1.4$\pm$0.8~sec.'' The reported half-life of 0.7(2)~s for $^{114}$Cs agrees with the currently accepted value of 570(20)~$\mu$s and the quoted half-life for $^{115}$Cs corresponds to the presently adopted value.

\subsection*{$^{116}$Cs}\vspace{0.0cm}
In the 1975 paper ``New Delayed-Proton Emitters $^{119}$Ba, $^{121}$Ba and $^{116}$Cs'' Bogdanov et al. reported the discovery of $^{116}$Cs \cite{1975Bog01}. The U-300 cyclotron of the Nuclear Reactions Laboratory at Dubna accelerated a $^{32}$S beam to a maximum energy of 190~MeV bombarding a target of natural zirconium enriched with $^{92}$Mo. ``In [the figure] we have shown the delayed-proton spectrum of $^{116}$Cs and the decay curve, from which it follows that T$_{1/2}$ = 3.9$\pm$0.4 sec.'' This half-life is included in the currently accepted average value of 3.85(13)~s for the isomeric state. Less than a month later Ravn et al. independently submitted their result on $^{116}$Cs measuring a half-life of 3.6(2)s \cite{1975Rav01}. The ground state was first observed two years later \cite{1977Bog02}.

\subsection*{$^{117}$Cs}\vspace{0.0cm}
Ravn et al. reported the first observation of $^{117}$Cs in the 1972 paper ``Very neutron-deficient isotopes of Cs studied by on-line isotope separator techniques'' \cite{1972Rav01}. Molten lanthanum was bombarded with 600 MeV protons from the CERN synchrocyclotron. $^{117}$Cs was formed in spallation reactions and identified with ISOLDE on-line isotope separator facility. ``By means of the on-line $\beta^+$ counter the mass chains down to 117 were accessible for decay measurements. Half-lives were determined by means of a least-squares analysis of the experimental data. The following results were obtained for the isotopes under study (the errors represent the estimated over-all uncertainties): $^{117}$Cs 8$\pm$2~s...'' This value agrees with the presently adopted half-lives of 8.4(6)~s.

\subsection*{$^{118-122}$Cs}\vspace{0.0cm}
The discovery of $^{118}$Cs, $^{119}$Cs, $^{120}$Cs, $^{121}$Cs, and $^{122}$Cs was reported in ``Identification of new neutron-deficient nuclides $^{76}$Rb and $^{118}$Cs. Half-lives of $^{78}$Rb, $^{119-124}$Cs, $^{126}$Cs'' by Chaumont et al. in 1969 \cite{1969Cha01}. 24 GeV protons from the CERN proton synchrotron bombarded tantalum targets. Rubidium and cesium ions were selectively emitted by surface ionization and separated with an on-line mass spectrometer. ``From the difference between the two spectra, the $^{118}$Cs peak appears to be statistically significant... The decay curves are analyzed with a least-squares program.'' The half-lives were listed in a table: 33(8)~s ($^{119}$Cs), 61.3(14)~s ($^{120}$Cs), 125.6(14)~s ($^{121}$Cs), 21.0(7)~s and 267(11)~s ($^{122}$Cs). $^{119}$Cs, $^{120}$Cs, and $^{121}$Cs have isomers with similar half-lives as the ground states and the reported values are consistent with either one. The value for $^{120}$Cs is included in the calculated average for the accepted ground state value of 61.3(11)~s for $^{120}$Cs. The reported half-lives for the ground and the isomeric states for $^{122}$Cs agree with the accepted values of 21.18(19)~s and 3.70(11)~min, respectively.

\subsection*{$^{123}$Cs}\vspace{0.0cm}
In 1954, Mathur and Hyde described the observation of $^{123}$Cs in ``Spectrometer studies of the radiations of neutron deficient isotopes of cesium and of the \emph{E3} isomers, Xe$^{127m}$ and Xe$^{125m}$'' \cite{1954Mat01}. The Berkeley 184-inch cyclotron was used to bombard calcium iodide with 130 MeV $\alpha$ particles. The resulting activities were measured with a G-M counter and a scintillation spectrograph following chemical separation. ``A new cesium activity of 6-minute half-life was produced along with 45-minute Cs$^{125}$ and 6.25-hour Cs$^{127}$.'' This half-life agrees with the currently accepted value of 5.87(4)~min.

\subsection*{$^{124}$Cs}\vspace{0.0cm}
The discovery of $^{124}$Cs was reported in ``Identification of new neutron-deficient nuclides $^{76}$Rb and $^{118}$Cs. Half-lives of $^{78}$Rb, $^{119-124}$Cs, $^{126}$Cs'' by Chaumont et al. in 1969 \cite{1969Cha01}. 24 GeV protons from the CERN proton synchrotron bombarded tantalum targets. Rubidium and cesium ions were selectively emitted by surface ionization and separated with an on-line mass spectrometer. ``For the $^{124}$Cs decay, in addition to the 26.5 s period, the computer analysis gives some indications of a much longer-lived activity in the order of minutes which would be compatible with the existence of a 12 min half-life mentioned by Alexander et al.'' The reported half-life of 26.5(15)~s is consistent with the presently adopted value of 30.9(4)~s. The quoted reference of Alexander et al. was published only as an internal report \cite{1967Ale01}.

\subsection*{$^{125}$Cs}\vspace{0.0cm}
``Mass assignments by isotope separation'' was published in 1954 by Michel and Templeton documenting the observation of $^{125}$Cs \cite{1954Mic01}. The Berkeley 184-inch cyclotron was used to bombard calcium iodide with 100 MeV $\alpha$ particles. The resulting activities were measured with a G-M counter and a scintillation spectrograph following chemical separation. ``A new isotope of cesium found by Mathur and Hyde of this laboratory, has been assigned to mass 125 and its half-life observed (from separated samples) to be 45$\pm$1 minutes.'' This half-life is included in the current calculated average value of 45(1)~min. The experimental details were described in the paper by Mathur and Hyde \cite{1954Mat01} mentioned in the quote. Mathur and Hyde submitted their paper less than a month after Michel and Templeton and give them credit for the measurement ``A G-M decay curve of the mass-125 fraction isolated with the help of Michel and Templeton in the time-of-flight mass separator showed a straight line decay of 45$\pm$1 minutes from an initial counting rate of 18 000 to less than 10 counts per minute.''

\subsection*{$^{126}$Cs}\vspace{0.0cm}
In the 1954 article ``New chain barium-126--cesium-126'' Kalkstein et al. announced the discovery of $^{126}$Cs \cite{1954Kal01}. Indium oxide was bombarded with a 140~MeV $^{14}$N beam produced by the Berkeley Crocker 60~inch cyclotron. $^{126}$Cs was detected following the decay of $^{126}$Ba which was produced in the fusion-evaporation reaction $^{115}$In($^{14}$N,3n). It was identified with a time-of-flight mass spectrograph and a NaI scintillation detector following chemical separation. ``Activity was found to collect only at the mass-126 position, and this decayed with a half-life of 1.6$\pm$0.2 minutes. The new chain is thus identified as Ba$^{126}$-Cs$^{126}$'' The quoted half-life agrees with the currently accepted half life of 1.64(2)~min.

\subsection*{$^{127}$Cs}\vspace{0.0cm}
Fink et al. observed $^{127}$Cs and published the results in their 1950 paper titled ``Neutron-deficient cesium isotopes'' \cite{1950Fin02}. The Berkeley 184-inch cyclotron was used to bombard $^{127}$I in the form of ammonium iodide with 60-MeV helium ions. The mass assignment of $^{127}$Cs was determined with a magnetic spectrograph following chemical separation. ``Mass spectrograph plates showing dark lines at mass numbers 127, 129, and 133 (stable carrier) were obtained, and transfer plates showed the radioactivity of lines 127 and 129; after the 5.5-hr. Cs$^{127}$ activity had effectively decayed out, the radioactive line at mass 129 still gave good transfer plates.'' This half-life agrees with the currently accepted value of 6.25(10)~h.

\subsection*{$^{128}$Cs}\vspace{0.0cm}
The 1951 paper ``A new short-lived isotope of cesium'' by Fink et al. documented the first observation of $^{128}$Cs \cite{1951Fin01}. Cesium chloride was bombarded with 96 MeV protons from the Rochester 130-inch cyclotron to produce $^{128}$Ba which decayed to $^{128}$Cs. Activities were measured following chemical separation. ``The best value for the half-life of Cs$^{128}$ is an average of 11 different determinations and is 3.13 $\pm$ 0.2 minutes.'' This value agrees with the currently accepted value of 3.640(14)~min.

\subsection*{$^{129}$Cs}\vspace{0.0cm}
Fink et al. observed $^{129}$Cs and published the results in their 1950 paper titled ``Neutron-deficient cesium isotopes'' \cite{1950Fin02}. The Berkeley 184-inch cyclotron was used to bombard $^{127}$I in the form of ammonium iodide with 60-MeV helium ions. The mass assignment of $^{129}$Cs was determined with a magnetic spectrograph following chemical separation. ``Mass spectrograph plates showing dark lines at mass numbers 127, 129, and 133 (stable carrier) were obtained, and transfer plates showed the radioactivity of lines 127 and 129; after the 5.5-hr. Cs$^{127}$ activity had effectively decayed out, the radioactive line at mass 129 still gave good transfer plates.'' The observed half-life of 31(1)~h agrees with the currently accepted value of 32.06(6)~h.

\subsection*{$^{130}$Cs}\vspace{0.0cm}
Smith et al. reported the discovery of $^{130}$Cs in their 1952 paper ``The disintegration of Cs$^{130}$'' \cite{1952Smi01}. The 23-MeV $\alpha$-particle beam of the Indiana University cyclotron bombarded an iodine target and $^{130}$Cs was formed in the reaction $^{127}$I($\alpha$,n). Activities curves were measured following chemical separation. ``The period of the resulting activity was measured repeatedly and was found to be 30$\pm$1 min.'' This half-life agrees with the currently accepted value of 29.21(4)~min. A previously reported 30~min half-life could not be uniquely be assigned to $^{130}$Cs \cite{1950Fin02}.

\subsection*{$^{131}$Cs}\vspace{0.0cm}
``Disintegration by consecutive orbital electron capture $_{56}$Ba$^{131}$ $\rightarrow$ $_{55}$Cs$^{131}$ $\rightarrow$ $_{54}$Xe$^{131}$'' was published in 1947 by Yu et al. documenting the observation of $^{131}$Cs \cite{1947Yu01}. $^{131}$Cs was formed by $\beta$-decay following neutron capture on barium and identified following chemical separation. ``The $_{55}$Cs$^{131}$ decays with a period of 10$\pm$0.3 days, emitting highly converted gamma-rays of 145$\pm$10 kev energy.'' This half-life agrees with the currently accepted value of 9.689(16)~d.

\subsection*{$^{132}$Cs}\vspace{0.0cm}
$^{132}$Cs was observed by Wapstra et al. and the results published in the 1953 paper ``Some radioactive isotopes of I, Xe and Cs'' \cite{1953Wap01}. 26 MeV deuterons from the Philips' synchrocyclotron in Amsterdam, Netherland were used to produce $^{132}$Cs in the reaction $^{133}$Cs(d,p2n). Gamma-ray spectra were measured with a scintillation spectrometer. ``The following results were obtained. 7.1 day $^{132}$Cs... The $\gamma$-ray in this isotope was found to have an energy of 685$\pm$10 keV,...'' It seems that Wapstra et al. did not claim discovery because they were aware of an internal report of the Plutonium Project \cite{1945Cam01,1950NBS01}. The value quoted for the half-life of $^{132}$Cs is in agreement with the currently accepted value of 6.479(7)~d.

\subsection*{$^{133}$Cs}\vspace{0.0cm}
The discovery of stable $^{133}$Cs was reported by Aston in the 1921 paper ``The constitution of the alkali metals'' \cite{1921Ast03}. The positive anode ray methods was used to identified $^{133}$Cs with the Cavendish mass spectrograph. ``The mass spectra obtained from c{\ae}sium (atomic weight 132$\cdot$81) have so far exhibited only one line, which when measured against the rubidium lines indicates a mass 133.''

\subsection*{$^{134}$Cs}\vspace{0.0cm}
Kalbfell and Cooley published the observation of $^{134}$Cs in their 1940 paper ``Radio-isotopes of Ba and Cs'' \cite{1940Kal01}. Cesium was irradiated with deuterons or neutrons from the 37-inch Berkeley cyclotron. Beta- and $\gamma$-ray spectra were measured following chemical separation. ``Cs bombarded with deuterons or neutrons consistently gave a 3-hour$\pm$10-minute period rather that the previously reported period of 1.5 hours... A long period (20$\pm$1 month) isotope, chemically identified as Cs and apparently isomeric with the 3-hour Cs$^{134}$, was prepared by neutron or deuteron bombardment of Cs.'' The observed half-lives of 3 hours and 20(1) months agree with the currently accepted values of 2.903(8)~h and 2.0648(10)~years. The quoted 1.5~h half-life refers to papers published in 1935 by Amaldi et al. (1.5~h) \cite{1935Ama01}, McLennen et al. (75~min) \cite{1935McL03}, and Latimer et al. (100~min) \cite{1935Lat01}. None of these paper cleanly identifies this half-life with $^{134}$Cs. In addition, Alexeeva \cite{1938Ale01} reported a cesium activity with a lower limit of 1~year.

\subsection*{$^{135}$Cs}\vspace{0.0cm}
In ``Characteristics of the fission product Cs$^{135}$'' the first measurement of $^{135}$Cs by Sugarman was reported in 1949 \cite{1949Sug02}. Samples of $^{135}$Xe were prepared from fission fragments from the Los Alamos homogenous pile. After the decay of the xenon activity, $\beta$-ray spectra were measured following chemical separation. ``Assignment of the new Cs activity to Cs$^{135}$ is very probable from the way in which it was prepared. This assignment is constant with the timing of the experiments in which the He gas-sweeping of the pile was retained for Xe adsorption after discarding the first half-hour sweeping to eliminate active gases from short-lived halogens. The only Cs isotopes which should be present in the decay products of the gas are Cs$^{133}$ arising from the decay of Xe$^{133}$ and Cs$^{135}$ from Xe $^{135}$. Since Cs$^{133}$ is a naturally occurring stable isotope, assignment of the activity to Cs$^{133}$ would mean that the expected mode of decay would be isomeric transition. The shape of the Al absorption curve of the Cs activity is characteristic of a $\beta^-$ spectrum rather than that of a conversion electron spectrum. The mass assignment of 135 appears most probable.'' The calculated half life of 2.1(7)$\times$10$^{6}$ years is consistent with the currently accepted value of 2.3(3)~My.

\subsection*{$^{136}$Cs}\vspace{0.0cm}

In the Plutonium Project paper ``Further study of the 13d Cs activity'' Glendenin reported the observation of $^{136}$Cs \cite{1951Gle03}. The Clinton Pile was used to irradiate UO$_3$ in order to produce $^{136}$Cs. Decay curves and $\gamma$-spectra were recorded following chemical separation. ``Finkle, Engelkemeir, and Sugarman \cite{1951Fin02} have shown that the 13d Cs is not produced in fission by a secondary neutron reaction, i.e., neutron capture by a cesium fission product with a mass number lower by 1, and also that it is not produced by neutron irradiation of xenon. They suggest that this activity is isomeric with Cs$^{135}$, Cs$^{137}$, or Cs$^{138}$ and is formed uniquely in fission as a primary product in low yield. Another possibility, which seems much more likely to the writer, is that the 13d Cs is the ``shielded'' nucleus Cs$^{136}$, which must be formed directly in fission (in low yield), since the indirect formation through a $\beta$-decay chain is prevented by the existence of stable Xe$^{136}$.'' The reported half-life agrees with the presently adopted value of 13.16(3)~d.

\subsection*{$^{137}$Cs}\vspace{0.0cm}

Turkevich et al. reported the identification of $^{137}$Cs in ``Mass assignment of 33y Cs$^{137}$'' as part of the Plutonium Project published in 1951 \cite{1951Tur01}. Xenon was irradiated with neutrons in the thimble of the Argonne Heavy-water Pile. Activities and absorption curves were measured following chemical separation. ``The identification of the cesium activity from neutron-irradiated xenon with the 33y Cs fission activity establishes the mass assignment of the activity as 137. The other possible mass of 135 which was allowed for the xenon-produced cesium activity had been previously eliminated for the fission-cesium activity.'' The reported half-life agrees with the presently adopted value of 30.08(9)~y. Townsend et al. had reported $\gamma$- and $\beta$-rays activities of $^{137}$Cs in 1948, however, they considered the isotope to be known, because they were aware of the results of the Plutonium Project \cite{1948Tow01}.

\subsection*{$^{138}$Cs}\vspace{0.0cm}
Seelmann-Eggebrecht identified $^{138}$Cs in their 1939 paper ``\"Uber einige aktive Xenon-Isotope'' \cite{1943See03}. Barium was irradiated with fast neutrons and cesium isotopes were produced in (n,p) reactions. Beta-decay curves were measured and a cesium isotope with a half-life of 33~min was detected: ``Sowohl nach seiner Halbwertszeit von 33 Minuten als auch nach der Absorptionskurve seiner $\beta$-Strahlen ist dieses mit dem bei der Uranspaltung nachgewiesenen 33-Minuten-C\"asium identisch. Nun hat W. Riezler das 3,8-Minuten-Xenon der Masse 137 zuordnen k\"onnen. Da dieses Xenon-Isotop jedoch keinen Folgek\"orper von 33 Minuten Halbswertzeit besitzt, mu\ss\ dieses C\"asium-Isotop der Masse 138 zugeordnet werden.'' (This isotope is according to the 33~min half-life as well as the $\beta$-ray absorption curve identical with the 33~min cesium isotope observed in uranium fission. W. Riezler was recently able to assign a 3.8~min xenon isotope to mass 137. The present cesium isotope must therefore be assigned to mass 138, because there is an isotope with a 33~min half-life in the decay chain of this xenon isotopes.) The 33~min half-life agrees with the currently adopted value of 33.41(18)~min. Half-lives of 35~min \cite{1939Hah05}, 30~min \cite{1939Gro01,1939Hey01}, 33~min \cite{1939Ate01,1939Hah03}, and 32.0(5)~min \cite{1940Gla02} had previous been reported in uranium fission without a unique mass assignment.

\subsection*{$^{139}$Cs}\vspace{0.0cm}
The identification of $^{139}$Cs was reported in 1939 by Heyn et al. in the article ``Transmutation of uranium and thorium by neutrons'' \cite{1939Hey01}. An uranyl nitrate solution was irradiated with a strong neutron source of the Philips X-Ray Laboratory at Eindhoven, Netherlands. Resulting activities were measured following chemical separation. ``C{\ae}sium, precipitated after 15 minutes with antimony chloride dissolved in hydrochloric acid, shower periods of 10 minutes, 30 minutes and a longer period... Based on the experiments described we suggest the following processes: $^{139}$Xe$\stackrel{(\sim0\cdot5m.)}{\longrightarrow}$ $^{139}$Cs$\stackrel{10m.}{\longrightarrow}$ $^{139}$Ba$\stackrel{87m.}{\longrightarrow}$ $^{139}$La (stable).'' The 10~min half-life assigned to $^{139}$Cs is in agreement with the currently accepted value of 9.37(5)~min. Only a week earlier Hahn and Strassmann reported the observation of a cesium activities with an upper limit of 8~min \cite{1939Hah05} which they assigned to $^{139}$Cs with a half-life of 6~min four months later \cite{1939Hah03}.

\subsection*{$^{140}$Cs}\vspace{0.0cm}
The paper ``Short-lived fission products. II. Cs$^{139}$ and Cs$^{140}$'' by Sugarman and Richter was published in 1950 reporting the first identification of $^{140}$Cs \cite{1950Sug01}. $^{140}$Cs was identified by timed precipitations of cesium silicotungstate precipitate from uranium. ``A plot of the normalized 12.8-day Ba$^{140}$ activity versus time of Cs precipitation is given in [the figure]. A least squares analysis of the data yielded 66$\pm$2 sec. for the half-life of Cs$^{140}$.'' This value is in agreement with the currently accepted value of 63.7(3)~s. Previously, Hahn and Strassman observed a 40~s cesium activity without a unique mass assignment \cite{1940Hah02}.

\subsection*{$^{141}$Cs}\vspace{0.0cm}
Wahl et al. reported the first identification of $^{141}$Cs in 1962 in the article ``Nuclear-charge distribution in low-energy fission'' \cite{1962Wah01}. $^{141}$Cs was produced from $^{235}$U fission induced by thermal neutrons. The neutrons were produced from reactions of 10 MeV deuterons accelerated by the Washington University cyclotron on a beryllium target. $^{141}$Cs was identified by timed separations of its daughters. ``The half-life value of Cs$^{141}$ obtained from the slope of the line is (25$\pm$3) sec. This value is in good agreement with the value of (24$\pm$2) sec determined by Fritze and Kennett.'' This value is consistent with the presently accepted value of 24.84(16)~s. The reference to Fritze and Kennett in the quote refers to a private communication.

\subsection*{$^{142,143}$Cs}\vspace{0.0cm}
``Half-lives of Cs$^{141}$, Cs$^{142}$, and Cs$^{143}$'' was published in 1962 by Fritze documenting the discovery of $^{142}$Cs and $^{143}$Cs \cite{1962Fri01}. The pneumatic-rabbit system of the McMaster Reactor was used to irradiate samples of U$^{235}$ in a flux of neutrons. Cesium was then precipitated out of the fission-product solution. ``[The figure] shows the resulting decay curve, indicating a half-life of 2.3$\pm$0.2 sec for Cs$^{142}$. This value is consistent with the upper limit of 8 sec given by Wahl et al. \cite{1962Wah01}. ... The decay curve of Cs$^{143}$ is shown in [the Figure], giving a half-life of 2.0$\pm$0.4 sec.'' The values quoted as the half-lives of $^{142}$Cs and $^{143}$Cs are consistent with the currently accepted values of 1.689(11)~s and 1.791(7)~s, respectively. As mentioned in the quote, earlier Wahl et al. was only able to extract an upper limit for the half-life of $^{142}$Cs \cite{1962Wah01}.

\subsection*{$^{144}$Cs}\vspace{0.0cm}
Amarel et al. observed $^{144}$Cs in 1967 as reported in the article ``Half life determination of some short-lived isotopes of Rb, Sr, Cs, Ba, La and identification of $^{93,94,95,96}$Rb as delayed neutron precursors by on-line mass-spectrometry'' \cite{1967Ama01}. $^{144}$Cs was produced by fission of $^{238}$U induced by 150 MeV protons from the Orsay synchrocyclotron. Isotopes were identified with a Nier-type mass spectrometer and half-lives were determined by $\beta$ counting. The measured half-life for $^{144}$Cs was listed in the main table with 1.06(10)~s which is consistent with the currently adopted value of 994(4)~ms.

\subsection*{$^{145,146}$Cs}\vspace{0.0cm}
$^{145}$Cs and $^{146}$Cs were first observed in 1971 by Tracy et al. in the article ``Half-lives of the new isotopes $^{99}$Rb, $^{98}$Sr, $^{145-146}$Cs'' \cite{1971Tra01}. Fission fragments from the bombardment of 50 MeV protons on $^{238}$U at the Grenoble cyclotron were studied. Beta-particles were measured at the end of an on-line mass spectrometer: ``The new isotopes $^{99}$Rb, $^{98}$Sr and $^{145-146}$Cs were observed and their half-life measured.'' The reported half-lives of 563(27)~ms and 189(11)~ms were listed in a table and are near the presently adopted values of 582(6)~ms and 323(6)~ms for $^{145}$Cs and $^{146}$Cs, respectively.

\subsection*{$^{147}$Cs}\vspace{0.0cm}
Wohn et al. reported the discovery of $^{147}$Cs in 1978 in their article ``Identification of $^{147}$Cs and half-life determinations for Cs and Ba isotopes with A=144-147 and Rb and Sr isotopes with A=96-98'' \cite{1978Woh01}. $^{147}$Ba was produced and identified by neutron induced fission of $^{235}$U at the On-line Separator f\"ur Thermisch Ionisierbare Spaltprodukte (OSTIS) facility of the Institut Laue-Langevin in Grenoble, France. ``Half-life determinations of Rb and Cs fission products available at an on-line mass separator have been made for several neutron-rich Rb, Sr, Cs, and Ba isotopes using both $\beta$-multiscale and $\gamma$-multispectra measurements. The half-lives and rms uncertainties (in sec) are... $^{147}$Cs, 0.218(009). This work provides an unambiguous identification of the new neutron-rich isotopes $^{147}$Cs.'' The observed value for the half-life is included in the calculation of the presently accepted value of 225(5)~ms.

\subsection*{$^{148}$Cs}\vspace{0.0cm}
Koglin et al. reported the identification of $^{148}$Cs in ``Half-lives of the new isotopes $^{100}$Rb, $^{100}$Sr, $^{148}$Cs and of $^{199}$Rb, $^{99}$Sr and $^{147}$Cs'' in 1978 \cite{1978Kog01}. $^{148}$Cs was produced and identified by neutron induced fission of $^{235}$U at the On-line Separator f\"{u}r Thermisch Ionisierbare Spaltprodukte (OSTIS) facility in Grenoble, France. ``An improvement of the ion source of the online fission product separator OSTIS allowed us to identify the new isotopes $^{100}$Rb (50$\pm$10~msec), $^{100}$Sr (170$\pm$80~ms), and $^{148}$Cs (130$\pm$40~ms).'' This half-life of $^{148}$Cs is included in the weighted average of the accepted value of 146(6)~ms.

\subsection*{$^{149-152}$Cs}\vspace{0.0cm}
``Experiments with intense secondary beams of radioactive ions'' was published in 1979 by Ravn reporting the observation of $^{149}$Cs, $^{150}$Cs, $^{151}$Cs, and $^{152}$Cs \cite{1979Rav01}. The CERN synchro-cyclotron accelerated protons to 600 MeV protons which were used to induce fission of $^{238}$U. The article represents an overview of the CERN ISOLDE program and the rates for neutron-rich cesium isotopes are shown in the figure. $^{149}$Cs, $^{150}$Cs, $^{151}$Cs, and $^{152}$Cs, were the heaviest isotopes measured which had not be reported previously with other production methods. No references to previous publications of these isotopes were given.

\section{Discovery of $^{117-153}$La}

Thirty-five lanthanum isotopes from A = 117--153 have been discovered so far; these include 2 stable, 19 neutron-deficient and 14 neutron-rich isotopes. According to the HFB-14 model \cite{2007Gor01,2007HFB01}, $^{182}$La should be the last odd-odd particle stable neutron-rich nucleus while the odd-even particle stable neutron-rich nuclei should continue through $^{191}$La. The proton dripline has already been reached at $^{117}$La. However, $^{118}$La and $^{199}$La have not been observed. In addition, it is anticipated that four more isotopes $^{113-116}$La could still have half-lives longer than 10$^{-9}$~ns \cite{2004Tho01}. Thus, about 40 isotopes have yet to be discovered corresponding to 53\% of all possible lanthanum isotopes.

\begin{figure}
	\centering
	\includegraphics[scale=.7]{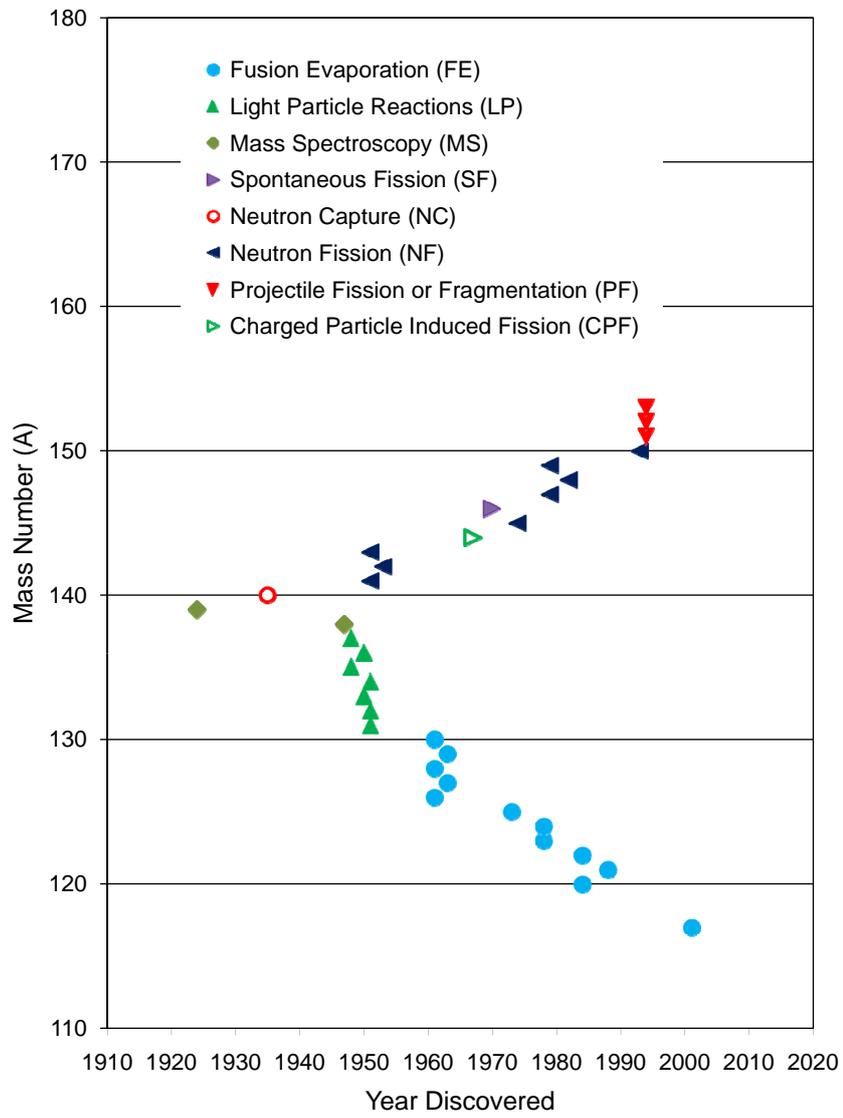}
	\caption{Lanthanum isotopes as a function of time when they were discovered. The different production methods are indicated.}
\label{f:year-la}
\end{figure}

Figure \ref{f:year-la} summarizes the year of first discovery for all lanthanum isotopes identified by the method of discovery. The range of isotopes predicted to exist is indicated on the right side of the figure. The radioactive lanthanum isotopes were produced using fusion evaporation reactions (FE), light-particle reactions (LP), neutron capture reactions (NC), neutron induced fission (NF), charged-particle induced fission (CPF), spontaneous fission (SF) and projectile fragmentation or fission (PF). The stable isotopes were identified using mass spectroscopy (MS). In the following, the discovery of each lanthanum isotope is discussed in detail.

\subsection*{$^{117}$La}\vspace{0.0cm}
``New strongly deformed proton emitter: $^{117}$La'' was published in 2001 by Soramel et al. documenting the first observation of $^{117}$La \cite{2001Sor01}. The reaction $^{64}$Zn($^{58}$Ni,p4n)$^{117}$La with a 310~MeV $^{58}$Ni beam from the Legnaro Tandem + LINAC accelerator was used and $^{117}$La was identified with a recoil mass spectrometer and a double sided silicon strip detector placed at the focal plane. ``The nucleus has two levels that decay to the ground state of $^{116}$Ba with E$_p$=783(6)~keV [T$_{1/2}$=22(5)~ms] and E$_p$=933(10)~keV [T$_{1/2}$=10(5)~ms].'' The quoted half-lives are in agreement with the currently accepted values of 23.5(26)~ms and 10(5)~ms. Less than six months later Mahud et al. \cite{2001Mah01} and most recently Liu et al. \cite{2011Liu01} confirmed the observation of the ground state but found no evidence for the isomeric state.

\subsection*{$^{120}$La}\vspace{0.0cm}
In the 1984 article ``Beta-delayed proton emission observed in new lanthanide isotopes'' Nitschke et al. reported the first observation of $^{120}$La \cite{1984Nit01}. A 253 MeV $^{64}$Zn beam from the Berkeley SuperHILAC was used to form $^{120}$La in the fusion-evaporation reaction $^{58}$Ni($^{64}$Zn,pn). Beta-delayed protons and characteristic X-rays were measured in coincidence at the on-line isotope separator OASIS. ``In an experiment with a $^{64}$Zn beam and a $^{58}$Ni target we observed beta-delayed protons in coincidence with Ba x-rays and conclude that the new precursor is $^{120}$La''. The measured half-life of 2.8(2)~s is the currently accepted value.

\subsection*{$^{121}$La}\vspace{0.0cm}
Sekine et al. published the observation of $^{121}$La in their 1988 paper ``Identification of a new isotope of $^{121}$La by means of element-selective mass separation'' \cite{1988Sek01}. An enriched $^{92}$Mo target was bombarded with a 4.7~MeV/u $^{32}$S beam from a tandem accelerator at the Japan Atomic Energy Research Institute. Gamma and beta singles as well as $\gamma-\beta$ coincidences were measured following on-line mass separation. ``Some of the short-lived $\gamma$ lines proved to coincide with Ba X-rays which are emitted in an electron-capture decay of $^{121}$La or in an internal conversion process associated with a transition in $^{121}$Ba, that is, the daughter of $^{121}$La. From these arguments we have assigned the observed 5.3-s activity to the $^{121}$La decay.'' This half-life is the currently accepted value.

\subsection*{$^{122}$La}\vspace{0.0cm}
In the 1984 article ``Beta-delayed proton emission observed in new lanthanide isotopes'' Nitschke et al. reported the first observation of $^{122}$La \cite{1984Nit01}. A 196 MeV $^{36}$Ar beam from the Berkeley SuperHILAC was used to form $^{122}$La in the fusion-evaporation reaction $^{92}$Mo($^{36}$Ar,$\alpha$pn). Beta-delayed protons and characteristic X-rays were measured in coincidence at the on-line isotope separator OASIS. ``A=122: Two experiments were performed at this mass value: one with a dual proton telescope where a total of about 1800 protons were recorded and one with the tape system where x- and $\gamma$-rays were measured in coincidence with protons. The x-ray spectrum in the second experiment showed only Ba K$_{\alpha}$- and K$_{\beta}$- lines which leads us to the conclusion that the new beta-delayed proton precursor is $^{122}$La.'' The observed half-life of 8.7(7)~s is in agreement with the currently accepted half-life of 8.6(5)~s.

\subsection*{$^{123,124}$La}\vspace{0.0cm}
The discovery of the isotopes $^{123}$La and $^{124}$La were first presented in the 1978 paper ``New neutron-deficient isotopes of lanthanum and cerium'' by Bogdanov et al. \cite{1978Bog01}. A 190 MeV $^{36}$S beam accelerated by the U-300 heavy-ion cyclotron of the Joint Institute for Nuclear Research (JINR) facility at Dubna, bombarded targets of $^{96}$Ru and $^{98}$Ru. The fusion-evaporation residues were mass separated with the on-line BEMS-2 facility and their X-ray and $\beta$ emission was detected with a Ge(Li) spectrometer and a plastic counter, respectively. Half-lives were determined from the X-ray decay curves. ``Seven isotopes $^{123-125}$La and $^{124-127}$Ce have been first observed and their half-lives and low-energy $\gamma$-ray data are reported.'' The observed half-life of $^{123}$La was 17(3)~s which corresponds to the currently accepted value. The observed half-life of 29(3)~s for $^{124}$La agrees with the currently accepted value of 29.21(17)~s. A 7.2(5)~min half-life quoted for $^{124}$La in the abstract of reference \cite{1963Pre01} is a typographical error; in the paper itself this half-life is assigned to $^{129}$La.

\subsection*{$^{125}$La}\vspace{0.0cm}
The first observation of $^{125}$La was published by Nakai et al. in their 1973 paper titled ``Prolate deformation in neutron-deficient lanthanum isotopes'' \cite{1973Nak01}. Fusion evaporation reactions of $^{14}$N on tin targets populated excited states in $^{125}$La. Gamma-ray angular distribution measurements and $\gamma$-$\gamma$ coincidences were recorded. ``The odd-A La isotopes from A = 125 to 137 have been studied by in-beam $\gamma$-ray spectroscopy of Sn($^{14}$N,xn)La reactions...'' Three transitions at 241, 437, and 604~keV were measured in $^{125}$La. In 1963 Preiss et al. \cite{1963Pre01} could only quote an upper limit for the lifetime of $^{125}$La. Bogdanov et al. who claimed the first observation of $^{125}$La in 1978 \cite{1978Bog01} was apparently not aware of the $\gamma$-ray spectroscopy work by Nakai et al.

\subsection*{$^{126}$La}\vspace{0.0cm}
In the 1961 paper, ``Experimental observation of a new region of nuclear deformation'' Sheline et al. described the observation of $^{126}$La \cite{1961She01}. $^{16}$O and $^{12}$C were accelerated by the Berkeley Hilac to 94~MeV and 117~MeV, respectively. $^{126}$La was formed in the fusion evaporation reactions $^{115}$In($^{16}$O,5n) and $^{121}$Sb($^{12}$C,7n) and identified in $\gamma$-ray measurements following chemical separation. ``Accordingly, we have produced the new nuclei La$^{126}$, La $^{128}$, and La$^{130}$ with half-lives of 1.0$\pm$0.3, 6.5 $\pm$1.0, and 9.0$\pm$1.0 min, respectively.'' The quoted value for the half-life of $^{126}$La is in accordance with the currently accepted value of 54(2)~s.

\subsection*{$^{127}$La}\vspace{0.0cm}
``Lanthanum isotopes in a possible new region of nuclear deformation`` was published in 1963 by Preiss et al. documenting the observation of $^{127}$La \cite{1963Pre01}. Indium foils were irradiated with either 66.4~MeV and 168~MeV $^{16}$O beams and $^{127}$La was produced in the fusion evaporation reactions $^{113-115}$In($^{16}$O,xn). Decay curves were recorded following chemical separation. ``Assignment of La$^{127}$ was made on the basis of the Ba$^{127}$ daughter decay and further confirmed using the Cs$^{127}$ granddaughter.'' The quoted half-life of 3.7(5)~min agrees with the currently accepted value for the isomeric state of 3.7(4)~min. The ground state was first observed ten years later \cite{1973Lei01}.

\subsection*{$^{128}$La}\vspace{0.0cm}
In the 1961 paper, ``Experimental observation of a new region of nuclear deformation'' Sheline et al. described the observation of $^{128}$La \cite{1961She01}. $^{16}$O was accelerated to 65 MeV and $^{12}$C to 84~MeV and 117~MeV by the Berkeley Hilac. $^{128}$La was formed in the fusion evaporation reactions $^{115}$In($^{16}$O,3n), $^{121}$Sb($^{12}$C,5n), and $^{123}$Sb($^{12}$C,7n) and identified in $\gamma$-ray measurements following chemical separation. ``Accordingly, we have produced the new nuclei La$^{126}$, La $^{128}$, and La$^{130}$ with half-lives of 1.0$\pm$0.3, 6.5 $\pm$1.0, and 9.0$\pm$1.0 min, respectively.'' The quoted value for the half-life of $^{128}$La is close to the currently accepted value of 5.18(14)~min.

\subsection*{$^{129}$La}\vspace{0.0cm}
``Lanthanum isotopes in a possible new region of nuclear deformation`` was published in 1963 by Preiss et al. documenting the observation of $^{129}$La \cite{1963Pre01}. Indium foils were irradiated with either 66.4~MeV and 168~MeV $^{16}$O beams and $^{129}$La was produced in the fusion evaporation reactions $^{113-115}$In($^{16}$O,xn). Decay curves were recorded following chemical separation. ``La$^{126}$ and La$^{129}$ have Ba daughters sufficiently similar in half-life to make their resolution difficult. The sum of the activities of these Ba daughters at the time of milking is plotted in [the Figure]. This curve can be resolved into 1.0- and 6.2-min components. The shorter half-life is assigned to La$^{126}$, it being the expectation that this more neutron-deficient odd-odd isotope has a shorter half-life than the odd-even La$^{129}$.'' The corrected half-life of 7.2(5)~min is close to the currently accepted value for the isomeric state of 11.6(2)~min. In the abstract this half-life was erroneously assigned to $^{124}$La. The half-life of the ground state was measured 16 years later \cite{1979Bro01}.

\subsection*{$^{130}$La}\vspace{0.0cm}
In the 1961 paper, ``Experimental observation of a new region of nuclear deformation'' Sheline et al. described the observation of $^{130}$La \cite{1961She01}. $^{12}$C was accelerated by the Berkeley Hilac to 53~MeV and 84~MeV. $^{130}$La was formed in the fusion evaporation reactions $^{121}$Sb($^{12}$C,3n) and $^{123}$Sb($^{12}$C,5n) and identified in $\gamma$-ray measurements following chemical separation. ``Accordingly, we have produced the new nuclei La$^{126}$, La $^{128}$, and La$^{130}$ with half-lives of 1.0$\pm$0.3, 6.5 $\pm$1.0, and 9.0$\pm$1.0 min, respectively.'' The quoted value for the half-life of $^{130}$La agrees with the currently accepted value of 8.7(1)min.

\subsection*{$^{131,132}$La}\vspace{0.0cm}
Gransden and Boyle published their observation of $^{131}$La and $^{132}$La in ``Neutron deficient isotopes of lanthanum'' in 1951 \cite{1951Gra01}. Protons of energies up to 90 MeV bombarded a barium target and $^{131}$La and $^{132}$La were isolated by a 180$^{\circ}$ mass spectrograph. ``Whereas the period of La$^{131}$ fits into the steadily decreasing series of half-lives for La isotopes of even neutron number, that of La$^{132}$ (4.5 hr) might be expected to be measured in minutes, in view of the corresponding series of known isotopes of odd neutron number.'' The measured half-life value for $^{131}$La of 58~min is in agreement with the currently accepted value of 59(2)~min. The quoted half-life value for $^{132}$La agrees with the currently accepted value of 4.8(2)~h.

\subsection*{$^{133}$La}\vspace{0.0cm}
The 1950 paper ``Mass spectrographic identification of radioactive lanthanum isotopes'' by Naumann et al. described the observation of $^{133}$La \cite{1950Nau01}. Cesium was bombarded with 50~MeV $\alpha$ particles and $^{133}$La was identified with a mass spectrograph and by measuring positrons, electrons, X- and $\gamma$-rays. ``The irradiation of cesium with 50-Mev alpha-particles resulted in a new lanthanum activity of 4.0-hour half life.'' This half-life is consistent with the currently accepted half life of 3.912(8)~h.

\subsection*{$^{134}$La}\vspace{0.0cm}
Stover reported the observation of $^{134}$La in the 1951 paper, ``New neutron-deficient radioactive isotopes of the light rare-earth region'' \cite{1951Sto01}. Lanthanum oxide was bombarded with 60- to 80~MeV protons and activities were measured with end-on type Geiger-M\"uller counters following chemical separation. ``In all bombardments with protons of 60- to 80- Mev energy, a 72.0-hr cerium activity remained after the other cerium activities had decayed out and their daughter activities had decayed or had been removed by chemical means. Chemical separation of the lanthanum daughter gave a 6.5~min activity.'' The measured half-life of 6.50(25)~min agrees with the currently accepted value of 6.45(16)~min.

\subsection*{$^{135}$La}\vspace{0.0cm}
$^{135}$La was discovered by Chubbuck and Perlman at Berkeley in the 1948 paper ``Neutron deficient isotopes of cerium and lanthanum'' \cite{1948Chu01}. The Berkeley 60-inch cyclotron was used to bombard CsNO$_3$ targets with 30~MeV $\alpha$ particles and $^{135}$La was produced in the reaction $^{133}$Cs($\alpha$,2n). $^{135}$La was identified with a $\beta$-ray spectrometer and by measuring X-rays, $\gamma$-rays and absorption spectra following chemical separation. ``The lanthanum fraction was separated and the decay curve resolved into two components of 2.1-hour and 19.5-hour half-lives... As indicated in [the Table], the 2.1-hour activity was formed in about 1/100 the yield of the 19.5-hour activity. With
the energy of helium ions employed (30 Mev), the ($\alpha$,2n) reaction is more prolific than the ($\alpha$,n) reaction. The order of the decay energies is also compatible with the assignment of the 2.1-hour period to La$^{136}$ and the 19.5-hour period to La$^{135}$.'' This half-life agrees with the presently adopted value of 19.5(2)~h. The assignment of the 2.1-h period to $^{136}$La was later shown to be incorrect \cite{1950Nau01}.

\subsection*{$^{136}$La}\vspace{0.0cm}
The 1950 paper ``Mass spectrographic identification of radioactive lanthanum isotopes'' by Naumann et al. described the observation of $^{136}$La \cite{1950Nau01}. Cesium was bombarded with $\alpha$ particles and $^{136}$La was identified with a mass spectrograph and by measuring positrons and X-rays. ``A search was made for La$^{136}$ and it proved to be the 10-minute period reported by Maurer from the irradiation of Barium with deuterons. The assignment was made to La$^{136}$ since the ratio of the 10-minute period to the 19-hour period increased with decreasing alpha-particle energy on cesium in the range below 30 Mev, as would be expected fit the ($\alpha$,n) reaction in relation to the ($\alpha$,2n) reaction.'' The measured half-life of 9.5~min. agrees with the currently accepted value of 9.87(3)~min. Maurer et al. - as referred to in the quote - could only assign the mass of the 10~min activity to either $^{136}$La, $^{137}$La, or $^{138}$La \cite{1947Mau01}. A previously reported half-life of 2.1~h \cite{1948Chu01} was evidently incorrect.

\subsection*{$^{137}$La}\vspace{0.0cm}
Inghram and Hess published ``The radioactive lanthanum and cerium isotopes of mass 137'' in 1948 documenting their observation of $^{137}$La \cite{1948Ing01}. CeO$_{2}$ was irradiated with neutrons in a graphite-moderated pile at Argonne. Resulting activities were then analyzed with a mass spectrograph. $^{137}$La was produced from lanthanum impurities in the sample. ``It is thus concluded that the isotope observed at mass 137 is lanthanum which has been formed by radioactive decay of Ce$^{137}$''. Previously, Chubbuck and Perlman could only determine a lower limit of $<$400~y for the half-life of $^{137}$La.

\subsection*{$^{138}$La}\vspace{0.0cm}
Inghram et al. discovered $^{138}$La in 1947 as reported in ``A new naturally occurring lanthanum isotope at mass 138'' \cite{1947Ing01}. A Nier-type mass spectrometer with a thermal ionization source was used to identify $^{138}$La. ``Peaks at masses 154, 155, 156, and 157 were observed. The 155, 156, and 157-peaks were present in relative abundances characteristic of La$^{139}$O$^{16}$, La$^{139}$O$^{17}$ and La$^{139}$O$^{18}$. The peak at 154 was unexpected but was always present to the same relative amount in four different lanthanum samples. The possibilities that this new peak might be due to Ce$^{138}$O$^{l6}$, Ba$^{138}$O$^{l6}$, Sm$^{154}$, or Gd$^{154}$ were ruled out because these elements would also have given other lines. We are obliged to attribute the 154 line to an isotope of Lanthanum, La$^{138}$, which is present to 0.089$\pm$0.002 percent.''

\subsection*{$^{139}$La}\vspace{0.0cm}
In 1924 Aston reported the first observation of stable $^{139}$La in ``Recent results obtained with the mass-spectrograph'' \cite{1924Ast04}. The mass spectra were measured at the Cavendish Laboratory in Cambridge, UK: ``Lanthanum (138.91) gives a single line of satisfactory strength at 139 and may therefore be taken as a simple element''.

\subsection*{$^{140}$La}\vspace{0.0cm}
$^{140}$La was observed by Marsh and Sugden published in the paper ``Artificial radioactivity of the rare earth elements'' \cite{1935Mar01}. Lanthanum oxide was irradiated with neutrons from a 400 mCi radon source in contact with powdered beryllium. ``On the other hand, we have found a considerable activity for lanthanum with a period of 2 days whilst Fermi reports that this element is inactive.'' The half-life of 1.9(2)~d quoted in a table is consistent with the presently accepted value of 1.6781(3)~d. The publication by Fermi mentioned in the quote refers to \cite{1935Ama01}.

\subsection*{$^{141}$La}\vspace{0.0cm}
Katcoff published the identification of $^{141}$La in an article of the Plutonium Project Record in 1951: ``Radiations from 3.7h La$^{141}$'' \cite{1951Kat01}. Uranyl nitrate was irradiated with slow neutrons produced with the Chicago cyclotron and the A=141 mass chain of the $^{141}$Ba-$^{141}$La-$^{141}$Ce relationship was established: ``About 75 min was then allowed for 3.7h La$^{141}$ to grow into the solution from its 18m Ba$^{141}$ parent... The $\beta$-decay curve shows a long-lived component (probably the 28d Ce$^{141}$ daughter of 3.7h La$^{141}$) and small amounts of 30~h and 1.5~h components; but the 3.7~h component greatly predominates.'' Hahn and Strassmann had originally reported this half-life in barium for the first time \cite{1942Hah01} modifying a previous observation of a single 14~min component \cite{1939Hah01} into two components of 18~min and 6~min. Hahn and Strassmann also observed the relationship of the 18~min half-life with a 3.5~h component in lanthanum. However, no specific mass assignment was made. In another paper in the Plutonium Project Record Goldstein mentioned the established relationship of the mass chain \cite{1951Gol01} referring to the work by Ballou and Burgus which was not included in the published record \cite{1943Bal01}.

\subsection*{$^{142}$La}\vspace{0.0cm}
Vanden Bosch reported the observation of $^{142}$La in the 1953 paper ``Radiations from $^{142}$Lanthanum'' \cite{1953Van01}. Barium was separated from uranium fission products and decay and absorption curves as well as $\gamma$-ray spectra were recorded following chemical separation of lanthanum from the barium. ``...the second barium precipitate yields only $^{141}$Lanthanum and $^{142}$Lanthanum with half lives of 3.6 hours and 77 minutes.'' The quoted value for $^{142}$La is consistent with the currently accepted value of 91.1(5)~min. In 1942 Hahn and Strassmann reported a lanthanum activity of 74(5)~min following the decay of a 6-min barium isotope \cite{1942Hah01}, however, they could not assign the mass number of this decay chain.

\subsection*{$^{143}$La}\vspace{0.0cm}
The first observation of $^{143}$La was reported in 1951 in the Plutonium Project paper ``Discovery of 19m La$^{143}$'' by Gest and Edwards \cite{1951Ges01}.
Uranyl nitrated was irradiated in the Clinton Pile and activities were measured following chemical separation. ``A study of the rate of growth of 33h Ce$^{143}$ in samples of active lanthanum isolated from uranium fission products has indicated that the parent La$^{143}$ has a half-life of about 19~min.'' This half-life is close to the presently adopted value of 14.2(1)~min. A 15~min half-life without a mass assignment was previously observed by Hahn and Strassmann \cite{1943Hah03}.

\subsection*{$^{144}$La}\vspace{0.0cm}
Amarel et al. observed $^{144}$La in 1967 as reported in their article ``Half life determination of some short-lived isotopes of Rb, Sr, Cs, Ba, La and identification of $^{93,94,95,96}$Rb as delayed neutron precursors by On-Line Mass-Spectrometry'' \cite{1967Ama01}. $^{144}$La was produced by fission of $^{238}$U induced by 150 MeV protons from the Orsay synchrocyclotron. Isotopes were identified with a Nier-type mass spectrometer and half-lives were determined by $\beta$ counting. The measured half-life for $^{144}$La was listed in the main table with 41(2)~s which is consistent with the currently adopted value of 40.8(4)~s.

\subsection*{$^{145}$La}\vspace{0.0cm}
$^{145}$La was identified in 1974 by Aronsson et al. in the paper ``Short-lived isotopes of lanthanum, cerium and praseodymium studied by SISAK-technique'' \cite{1974Aro01}. A uranium target was irradiated with 14~MeV neutrons and after chemical separation $^{145}$La was identified by measuring $\gamma$-ray spectra with a Ge(Li)-detector system. ``So far, no conclusive data have been presented for the isotopes $^{145}$La and $^{146}$La. However, information available from other laboratories along with our own data suggests that the $\gamma$-groups with half-lives 20 $\pm$ 5 sec and 11 $\pm$ 1 sec should be associated with the decay of $^{145}$La and $^{146}$La, respectively'' The quoted half-life of $^{145}$La is consistent with the currently adopted value of 24.8(20)~s. The information from other laboratories mentioned in the quote refers to a conference proceeding, an unpublished report and a private communication. The observation of a 100~keV $\gamma$-ray in the spontaneous fission of $^{252}$Cf \cite{1970Wat01,1971Hop01} was evidently incorrect \cite{1999Zhu01}.

\subsection*{$^{146}$La}\vspace{0.0cm}

In ``A Study of the Low-Energy Transitions Arising from the Prompt De-Excitation of Fission Fragments,'' published in 1970, Watson et al.\ reported the first observation of $^{146}$La \cite{1970Wat01}. Fission fragments from the spontaneous fission of $^{252}$Cf were measured in coincidence with X-rays and conversion electrons. In a table two $\gamma$-ray transitions with energies of 64~keV and 131~keV were listed with confidence level A.

\subsection*{$^{147}$La}\vspace{0.0cm}
Engler et al. observed $^{147}$La as reported in the 1979 article ``Half-life measurements of Rb, Sr, Y, Cs, Ba, La and Ce isotopes with A=91$-$98 and A=142$-$149'' \cite{1979Eng01}. A $^{235}$U target was irradiated by thermal neutrons at the Soreq Nuclear Research Centre in Yavne, Israel. $^{148}$La was identified with the Soreq-On-Line-Isotope-Separator (SOLIS). ``Half-lives of the extremely neutron-rich isotopes $^{91-98}$Rb, $^{94-98}$Sr, $^{96-98}$Y, $^{142-146}$Cs, $^{143-148}$Ba, $^{144-149}$La, and $^{149}$Ce were determined.'' The half-life determined for $^{147}$La was 4.4(5)~s. This value agrees with the currently accepted value of 4.015(8)s. The observation of 58.1~keV and 158.9~keV $\gamma$-rays in the spontaneous fission of $^{252}$Cf \cite{1971Hop01} was evidently incorrect \cite{1999Zhu01}.

\subsection*{$^{148}$La}\vspace{0.0cm}
The 1982 paper ``P$_{\emph{n}}$-values of short-lived Sr, Y, Ba, and La precursors'' by Gabelmann et al. discussed the observation of $^{148}$La \cite{1982Gab01}. Fission of $^{235}$U by thermal neutrons was measured at the OSTIS facility in Grenoble and $^{148}$La was identified by neutron multiscaling and $\gamma$-ray multispectra measurements. The half-life of $^{148}$La is given in a table as 1550(30)~ms. This value is close to the presently accepted value of 1.26(8)~s.

\subsection*{$^{149}$La}\vspace{0.0cm}
Engler et al. observed $^{149}$La for the first time as reported in the 1979 article ``Half-life measurements of Rb, Sr, Y, Cs, Ba, La and Ce isotopes with A=91-98 and A=142-149'' \cite{1979Eng01}. A $^{235}$U target was irradiated to thermal neutrons at the Soreq Nuclear Research Centre in Yavne, Israel. $^{149}$La was identified with the Soreq-On-Line-Isotope-Separator (SOLIS). ``The isotopes $^{147,148}$Ba and $^{149}$La were identified for the first time and their half-lives measured. The values obtained, in seconds, are 0.72$\pm$0.07 for $^{147}$Ba, 0.47$\pm$0.20 for $^{148}$Ba and 1.2$\pm$0.4 for $^{149}$La.'' The quoted half-life for $^{149}$La agrees with the currently accepted value of 1.05(3)~s.

\subsection*{$^{150}$La}\vspace{0.0cm}
In the 1993 article ``Delayed-neutron branching ratios of precursors in the fission product region'' Rudstam et al. reported the observation of $^{150}$La \cite{1993Rud01}. $^{150}$La was produced and identified at the OSIRIS isotope-separator on-line facility at the Studsvik Neutron Research Laboratory in Nyk\"oping, Sweden. ``The 0.856-s neutron activity at mass number 150 is attributed to the hitherto unknown nuclide $^{150}$La.'' While the ENSDF data base lists the measured half-life of 0.86(5)~s as the accepted value, the NUBASE evaluation adopts the more recent 510(30)~ms half-life measured by Okano et al.\ \cite{1995Oka01}.

\subsection*{$^{151-153}$La}\vspace{0.0cm}
In 1994, Bernas et al. published the discovery of $^{151}$La, $^{152}$La, and $^{153}$La in ``Projectile fission at relativistic velocities: a novel and powerful source of neutron-rich isotopes well suited for in-flight isotopic separation'' 1994 \cite{1994Ber01}. The isotopes were produced using projectile fission of $^{238}$U at 750 MeV/nucleon on a lead target at GSI, Germany. ``Forward emitted fragments from $^{80}$Zn up to $^{155}$Ce were analyzed with the Fragment Separator (FRS) and unambiguously identified by their energy-loss and time-of-flight.'' This experiment yielded 106, 20, and 5 counts of $^{151}$La, $^{152}$La, and $^{153}$La, respectively.

\section{Discovery of $^{121-154}$Pr}

Thirty-two praseodymium isotopes from A = 121--154 have been discovered so far; these include 1 stable, 18 neutron-deficient and 13 neutron-rich isotopes.
According to the HFB-14 model \cite{2007Gor01,2007HFB01}, $^{184}$Pr should be the last odd-odd particle stable neutron-rich nucleus while the odd-even particle stable neutron-rich nuclei should continue through $^{195}$Pr. The proton dripline has already been crossed with the observation of the proton emitter $^{121}$Pr. However, $^{122}$Pr and $^{123}$Pr have not been observed. In addition, it is anticipated that six more isotopes $^{115-120}$Pr could still have half-lives longer than 10$^{-9}$~ns \cite{2004Tho01}.  Thus, about 42 isotopes have yet to be discovered corresponding to 55\% of all possible praseodymium isotopes.

\begin{figure}
	\centering
	\includegraphics[scale=.7]{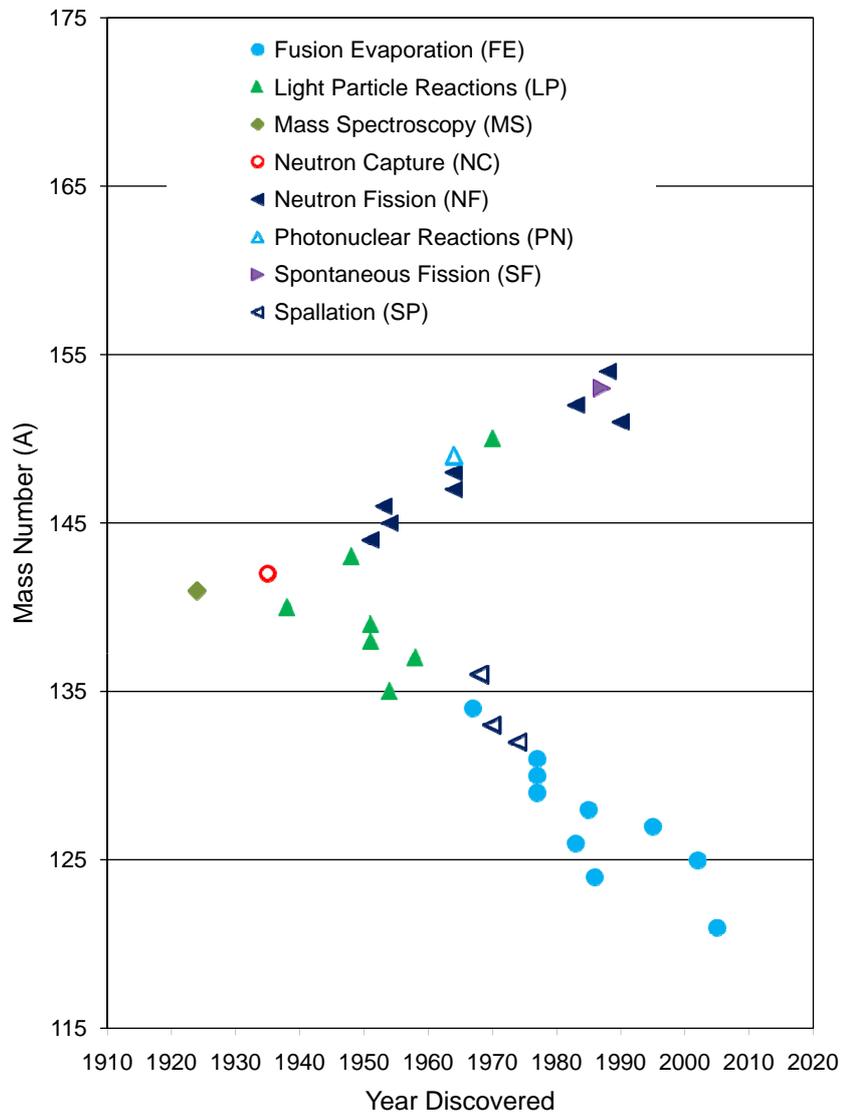}
	\caption{Praseodymium isotopes as a function of time when they were discovered. The different production methods are indicated.}
\label{f:year-pr}
\end{figure}

Figure \ref{f:year-pr} summarizes the year of first discovery for all praseodymium isotopes identified by the method of discovery. The range of isotopes predicted to exist is indicated on the right side of the figure. The radioactive praseodymium isotopes were produced using fusion evaporation reactions (FE), light-particle reactions (LP), photo-nuclear reactions (PN), neutron induced fission (NF), spontaneous fission (SF), neutron capture (NC), and spallation reactions (SP). The stable isotope was identified using mass spectroscopy (MS). In the following, the discovery of each praseodymium isotope is discussed in detail.

\subsection*{$^{121}$Pr}\vspace{0.0cm}
$^{121}$Pr was first observed in 2005 by Robinson et al. with the results published in their paper ``Ground state proton radioactivity from $^{121}$Pr: When was this exotic nuclear decay mode first discovered?'' \cite{2005Rob01}. A $^{92}$Mo target was bombarded with a 240~MeV $^{36}$Ar beam from the ATLAS accelerator facility at Argonne. $^{121}$Pr was formed in the fusion-evaporation reaction $^{92}$Mo($^{36}$Ar,1p6n) and identified at the end of the Argonne Fragment Mass Analyzer. ``Ground-state proton radioactivity has been identified from $^{121}$Pr. A transition with a proton energy of E$_p$ = 882(10)~keV [Q$_p$ = 900(10)~keV] and half-life t$_{1/2}$ = 10$^{+6}_{-3}$~ms has been observed and is assigned to the decay of a highly prolate deformed 3/2$^+$ or 3/2$^-$ Nilsson state.'' This half-life corresponds to the currently accepted value. A previously reported half-life of 1~s \cite{1973Bog02} was evidently incorrect.

\subsection*{$^{124}$Pr}\vspace{0.0cm}
The first observation of $^{124}$Pr was reported by Wilmarth et al. in their 1986 paper entitled ``Beta-delayed proton emission in the lanthanide region'' \cite{1986Wil01}. $^{124}$Pr was produced in heavy-ion induced-fusion evaporation reactions at the Berkeley Super HILAC. Beta-delayed particles, X-rays and $\gamma$-rays were measured following mass separation with the on-line isotope separator OASIS. ``A $\beta$-delayed proton activity with a half-life of 1.2$\pm$0.2 s was assigned to the new isotope $^{124}$Pr on the basis of the Ce x-rays observed in coincidence with the protons.'' This half-life corresponds to the currently accepted value.

\subsection*{$^{125}$Pr}\vspace{0.0cm}
``Observation of excited states in the near-drip-line nucleus $^{125}$Pr'' was published in 2002 by Wilson et al. describing the identification of $^{125}$Pr \cite{2002Wil01}. A 240 MeV $^{64}$Zn from the ATLAS accelerator at Argonne bombarded a zinc target and $^{125}$Pr was formed in the fusion-evaporation reaction $^{64}$Zn($^{64}$Zn,p2n). Evaporation residues were detected with the Fragment Mass Analyzer (FMA), charged particles, neutrons and $\gamma$ rays were measured with the Microball, the Neutron Shell, and Gammasphere, respectively. ``The isotope $^{125}$Pr has previously been identified in a mass-separation experiment by its $\beta^+$ decay to $^{125}$Ce \cite{1995Osa01}; the half-life of the ground state was measured as 3.3~s, but no excited states or $\gamma$-ray decays were observed. In the present work, $\gamma$-ray transitions have been observed to high spins in $^{125}$Pr.'' Reference \cite{1995Osa01} was a conference proceeding and the result were never published in the refereed literature.

\subsection*{$^{126}$Pr}\vspace{0.0cm}
Nitschke et al. discovered $^{126}$Pr in the 1983 paper ``New beta-delayed proton emitter in the lanthanide region'' \cite{1983Nit01}. A $^{40}$Ca beam from the Berkeley Super HILAC bombarded a gas-cooled $^{92}$Mo target. $^{126}$Pr was identified with the on-line isotope separator OASIS. ``The predicted half-lives are 2~s for $^{126}$Nd and 3~s for $^{126}$Pr. From the longer half-life and much larger cross section, it can be concluded that the observed activity is most likely $^{126}$Pr''. The observed half-life of 3.2(6)~s agrees with the currently adopted value of 3.12(18)~s.

\subsection*{$^{127}$Pr}\vspace{0.0cm}
Gizon et al. published the observation of $^{127}$Pr in their 1994 paper titled ``$^{127}$Ce levels populated in the 4.2~s $^{127}$Pr beta-decay'' \cite{1995Giz01}. A 210 MeV $^{40}$Ca beam from the Grenoble SARA accelerator bombarded an enriched metallic $^{92}$Mo target. $^{127}$Pr was formed in the fusion evaporation reaction $^{92}$Mo($^{40}$Ca,p$\alpha$) and identified using on-line mass separation with the SARA/IGISOL technique. ``From new measurements, a half-life of 4.2$\pm$0.3~s has been clearly established in $^{127}$Pr. This value confirms our first result and rules out other data.'' This half-life corresponds to the currently accepted value. The first results mentioned in the quote were only published in a proceeding and the disputed previously measured half-life was 7.7(6)~s \cite{1994Sek01}.

\subsection*{$^{128}$Pr}\vspace{0.0cm}
$^{128}$Pr was discovered in 1985 by Wilmarth et al. and the results were published in the paper entitled ``Beta-delayed proton precursors with 59$\leqslant$ Z $\leqslant$ 62'' \cite{1985Wil01}. A 170~MeV $^{40}$Ca beam from the Berkeley Super HILAC was used to produce $^{128}$Pr in the fusion-evaporation reaction $^{92}$Mo($^{40}$Ca,3pn). Charged-particles, X-rays and $\gamma$-rays were measured following mass separation with the on-line separator OASIS. ``A weak $\beta$-delayed proton activity with a half-life of 3.2$^{+0.5}_{-0.5}$s observed in coincidence with Cs K x-rays can now be unambiguously assigned to the new isotope $^{128}$Pr.'' The quoted half-life agrees with the currently accepted value of 2.84(9)~s.

\subsection*{$^{129-131}$Pr}\vspace{0.0cm}
The first identification of $^{129}$Pr, $^{130}$Pr, and $^{131}$Pr, was reported in 1977 by Bogdanov et al. in ``New neutron-deficient isotopes of barium and rare-earth elements'' \cite{1977Bog01}. The Dubna U-300 Heavy Ion Cyclotron accelerated a $^{32}$S beam which bombarded enriched targets of $^{102}$Pd and $^{106}$Cd. The isotopes were identified with the BEMS-2 isotope separator. ``In the present paper, isotopes were mainly identified by measuring the $\gamma$-ray and X-ray spectra of the daughter nuclei formed as a result of measuring the $\beta^+$ decay.'' The reported half-lives of 24(5)~s for $^{129}$Pr, 28(6)~s for $^{130}$Pr, and 100(20)~s for $^{131}$Pr are close to the accepted values of 30(4)s, 40(4)s, and 1.50(3)~min, respectively.

\subsection*{$^{132}$Pr}\vspace{0.0cm}
``Method for obtaining separated short-lived isotopes of rare earth elements'' was published in 1974 by Latuszynski et al. documenting their observation of $^{132}$Pr \cite{1974Lat02}. A 0.05~mm thick tantalum target was bombarded with 660~MeV protons from the JINR synchrocyclotron in Dubna, Russia. Gamma-ray spectra and decay curves were measured at the end of an electromagnetic separator. ``Using the method proposed for investigations in the field of nuclear spectroscopy the gamma-spectra of short-living isotopes with T$_{1/2} \le$ 1 minute have been measured. The new isotopes $^{161}$Yb (4.2~min), $^{148}$Dy (3.5~min) $^{132}$Pr (1.6~min) have been identified.'' The observed half-life is in agreement with the currently accepted half-life of 1.49(11)~min. An English translation of the work is published in reference \cite{1974Lat01}. The publication by Arlt et al. \cite{1974Arl01} which shares one common co-author (Latuszynski) reporting the same half-life (1.6(3)~min) seems to originate from the same experiment.

\subsection*{$^{133}$Pr}\vspace{0.0cm}
$^{133}$Pr was discovered by Abdurazakov et al. in the 1970 paper ``New isotopes $^{133}$Pr, $^{134}$Nd, and $^{135}$Nd; decay schemes of $^{134}$Pr and $^{135}$Pr'' \cite{1970Abd01}. Spallation reactions were induced by 660 MeV protons irradiating a gadolinium target at the Joint Institute for Nuclear Reactions in Dubna, Russia. The reaction products were then chemically separated and identified by the measured $\gamma$-ray spectra. ``Thus, the arguments presented above permit us to infer that the observed new 7$\pm$3 min activity belongs to $^{133}$Pr, a previously unknown isotope whose decay leads to the emission of 133~keV $\gamma$ rays.'' This half-life is in agreement with the currently accepted value of 6.5(3)~min.

\subsection*{$^{134}$Pr}\vspace{0.0cm}
$^{134}$Pr was correctly identfied in 1967 by Clarkson et al. as reported in ``Collective excitations in neutron-deficient barium, xenon, and cerium isotopes \cite{1967Cla01}. A 90~MeV $^{12}$C beam from the Berkeley HILAC bombarded a copper iodide target to form $^{134}$Pr in the fusion-evaporation reaction $^{127}$I($^{12}$C,5n). Gamma-ray spectra were measured following chemical separation. ``The decay of $^{134}$Pr to $^{134}$Ce, with a half-life of 17$\pm$2 min, was also studied''. This half-life corresponds to the currently accepted value. The ENSDF database lists this half-life as the ground state while in the NUBASE evaluation it is listed as the isomeric state. The corresponding second state with a half-life of $\sim$11~min was first observed six years later by Arlt et al.\ in a paper titled ``A New Isomeric State in $^{134}$Pr and Excited States of $^{134}$Ce'' \cite{1973Arl01}. Previously reported half-lives of 1~h \cite{1960Lav01} and 40(7)~min \cite{1963Lav01} was evidently incorrect.

\subsection*{$^{135}$Pr}\vspace{0.0cm}
``Neutron-deficient activities of praseodymium'' was published in 1954 by Handley and Olson documenting their observation of $^{135}$Pr \cite{1954Han03}. An enriched $^{136}$Ce target was bombarded with 22.4~MeV protons from the Oak Ridge 86-inch cyclotron. Beta-decay curves and $\gamma$-ray spectra were measured following chemical separation. ``In another bombardment of enriched Ce$^{136}$ with 22.4-Mev protons the praseodymium fraction from an initial separation was permitted to decay; then, after a second separation, 22-hr Ce$^{135}$ was found to be present in the cerium fraction. Thus, the 22-min activity is the parent of Ce$^{135}$ and is assigned to Pr$^{135}$.'' This half-life is consistent with the currently accepted half-life of 24(2)~min.

\subsection*{$^{136}$Pr}\vspace{0.0cm}
The assignment of $^{136}$Pr was reported in ``A new neodymium isotope (A = 136) and its decay properties'' by Zhelev et al. \cite{1968Zhe01}. A 660 MeV proton beam bombarded a gadolinium target and $^{136}$Pr was identified with $\gamma$- and $\beta$-ray spectra following chemical separation. ``Here we show that the $\beta^+$ and $\gamma$ radiations observed in the 55-min neodymium activity are not due to decay of $^{137}$Nd but to decay of a new isotope, $^{136}$Nd, and its daughter $^{136}$Pr.'' The observed half-life of 13.5~min for $^{136}$Pr is agrees with the currently accepted value of 13.1(1)~min. Previously reported half-lives of 70~min \cite{1954Han03} and 1.00(15)~h \cite{1958Dan01} was evidently incorrect.

\subsection*{$^{137}$Pr}\vspace{0.0cm}
``Discovery of Pr$^{137}$'' was published in 1958 by Dahlstrom et al. documenting their observation of $^{137}$Pr \cite{1958Dah01}. Cerium oxide targets were irradiated with 20- to 40-MeV protons. Decay curves were measured with an end-on Geiger-M\"uller tube after chemical and mass separation in a Nier type spectrograph. ``In the course of a general study of neutron-deficient isotopes of praseodymium formed under proton bombardment of cerium oxide, the extracted Pr portion of a target exposed at 40 Mev clearly showed Pr$^{137}$ in a mass spectrograph. Thus separated, this isotope repeatedly showed an easily measurable half-life of 1.5~hours.'' The quoted half-life of 1.5(1)~h is in agreement with the currently accepted value of 1.28(3)~h. Previously determined limits of $<$4~min or $>$1~y \cite{1954Han03} were evidently incorrect. The same group simultaneously submitted another paper with a different first author confirming the results \cite{1958Dan01}. It was published in the same volume of the Canadian Journal of Physics immediately following the paper by Dahlstrom et al.

\subsection*{$^{138-139}$Pr}\vspace{0.0cm}
Stover reported the observation of $^{138}$Pr and $^{139}$Pr in the 1951 paper, ``New neutron-deficient radioactive isotopes of the light rare-earth region'' \cite{1950Sto01}. Cerium metal was bombarded with 10-, 20-, and 32-Mev protons and activities were measured with end-on type Geiger-M\"uller counters following chemical separation.  ``In the bombardments with 32-Mev protons a 120-min praseodymium activity appeared in higher yield than the 4.50-hr Pr$^{139}$. On the basis of the approximate threshold of 30 Mev, it was allocated to Pr$^{138}$ as the product of a (p,3n) reaction on Ce$^{140}$... The 4.50-hr praseodymium activity was formed in bombardments of 20- and 32-Mev protons on cerium but did not appear at 10 Mev. It was thus assigned to Pr$^{139}$ as the product of a (p,2n) reaction on Ce$^{140}$. Confirmation of the mass assignment was made by showing it to be the parent of the 140-day Ce$^{139}$''. The observed half-lives of 120(5)~min for $^{138}$Pr and 4.50(2)~h for $^{139}$Pr agree with the presently adopted values of 2.12(4)~h and 4.41(4)~h, respectively. The half-life for $^{138}$Pr corresponds to an isomeric state. The ground state was first observed 15 years later \cite{1966Gro01}.

\subsection*{$^{140}$Pr}\vspace{0.0cm}
The first detection of $^{140}$Pr was reported in 1938 by Pool and Quill in ``Radioactivity induced in the rare earth elements by fast neutrons'' \cite{1938Poo02}. Fast and slow neutrons were produced with 6.3 MeV deuterons from the University of Michigan cyclotron. Decay curves were measured with a Wulf string electrometer. ``With fast neutrons a very strong 3.5 min period was observed which is by far the strongest positron period in the rare earth group of elements... since praseodymium has only one stable isotope, the reaction equations for the short period may be written: $_{59}$Pr$^{141}$ + $_{0}$n$^{1}$ $\rightarrow$ $_{59}$Pr$^{140}$ + 2$_{0}$n$^{1}$; $_{59}$Pr$^{140}$ $\rightarrow$ $_{58}$Ce$^{140}$ + $_{1}$e$^{0}$ (3.5 min).'' The observed half-life of 3.5 min is in agreement with the currently accepted value of 3.39(1)~min.

\subsection*{$^{141}$Pr}\vspace{0.0cm}
In 1924 Aston reported the first observation of stable $^{141}$Pr in ``Recent results obtained with the mass-spectrograph'' \cite{1924Ast04}. The mass spectra were measured at the Cavendish Laboratory in Cambridge, UK: ``A commercial sample of praseodymium (140.92) showed the same strong line but with indication of one at 141, so the experiment was repeated with a highly-purified sample prepared by Auer von Welsbach. This gave only one line at 141, indicating that praseodymium is most probably simple.''

\subsection*{$^{142}$Pr}\vspace{0.0cm}
$^{142}$Pr was identified by Marsh and Sugden and published in the paper ``Artificial radioactivity of the rare earth elements'' \cite{1935Mar01}. Praseodymium oxide was irradiated with neutrons from a 400 mCi radon source in contact with powdered beryllium. ``We have confirmed the 19 hour period of praseodymium and the 40 minute period of samarium discovered by Fermi and his co-workers...'' The half-life of 19.0(5)~h quoted in a table is consistent with the presently accepted value of 19.12(4)~h. Fermi and his co-workers mentioned in the quote did report a 18~h half-life but did not assign it to a specific nuclide \cite{1935Ama01}.

\subsection*{$^{143}$Pr}\vspace{0.0cm}
The paper ``Radioactive cerium and praseodymium'' by Pool and Krisberg from 1948 reported the discovery of $^{143}$Pr \cite{1948Poo01}. $^{143}$Pr was produced through the bombardment of cerium with 10~MeV deuterons from the Ohio State 42-inch cyclotron. Activity and absorption measurements were made with a Wulf unifilar electrometer. ``In accordance with the reactions given above, the 13.5-day period of praseodymium is assigned Pr$^{143}$ which decays with the emission of a 0.83-Mev $\beta$-particle to stable Nd$^{143}$.'' The measured half-life of 13.5(1)~d agrees with the currently adopted value of 13.57(2)d. A 13.0(5)~d half-life had been reported earlier without a mass assignment \cite{1944Jol01}. A 13.8~d half-life for $^{143}$Pr was independently discovered during the Plutonium Project but was only published in the unclassified literature in 1951 \cite{1951Kat05,1951Bal02,1951Bal03,1951Bal04}.

\subsection*{$^{144}$Pr}\vspace{0.0cm}
$^{144}$Pr was identified by Burgus et al. as part of the Plutonium Project ``Characteristics of the 275d $^{144}$Ce'' \cite{1951Bur01}. Decay curves were measured following chemical separation of fission products from thermal neutron radiation of $^{235}$U. ``The 275d Ce has a mass of 144 and decays to the 17.5m Pr. It has a $\beta$ component of 0.35~MeV energy, but it has no $\gamma$ ray.'' This half-life agrees with the presently adopted value of 7.2(3)~min. In another paper of the Plutonium Project Newton et al. claimed the discovery of 17.5m $^{144}$Pr, however, no mass identification was made and the authors relied on the mass determination of the long-lived cerium which at the time of the original article was assumed to be 143 \cite{1951New01}.

\subsection*{$^{145}$Pr}\vspace{0.0cm}
$^{145}$Pr was discovered by Markowitz et al. in 1954: ``A new 3.0min Ce fission product and its 5.95-hr Pr daughter'' \cite{1954Mar01}. Uranyl nitrate was irradiated with neutrons in the Brookhaven pile and the cerium and praseodymium were chemically separated. ``The decay of this sample was followed with an end-window proportional counter and the Pr$^{145}$ activity was observed to decay exponentially through 14 half-lives with a half-time of 5.93 hours. A weighted average gives a value of 5.95 hours with an estimated maximum error of 0.10 hr. Several other decay curves, which were not determined as carefully as these two, also yielded values very close to 6.0 hours.'' The observed half-life of 5.95(10)~h is in agreement with the currently accepted value of 5.984(10)~h. A 4.5~h half-life of praseodymium was reported in other papers of the Plutonium Project, however, no firm mass assignments were made \cite{1951Bal01,1951Kat04}.

\subsection*{$^{146}$Pr}\vspace{0.0cm}
$^{146}$Pr was identified in 1953 by Caretto and Katcoff in the paper entitled ``Short-lived cerium isotopes from uranium fission'' \cite{1953Car01}. $^{146}$Pr was produced from irradiation of uranyl nitrate and decay curves were measured with an end-window Geiger tube. ``The half-life of Pr$^{146}$ was found to be 24.4$\pm$0.5 minutes as determined from twelve decay curves, each followed for four to nine half-lives.'' This half-life agrees with the presently accepted value of 24.15(18)~min. Caretto and Katcoff did not consider their measurement a new discovery quoting an unpublished report by Schuman \cite{1945Sch01} and a paper by H. G\"otte \cite{1946Got01} who measured a half-life of 25~min but did not make a mass assignment.

\subsection*{$^{147-149}$Pr}\vspace{0.0cm}
$^{147}$Pr, $^{148}$Pr and $^{149}$Pr were discovered by Hofman and Daniels in 1964: ``Some short-lived isotopes of cerium and praseodymium'' \cite{1964Hof01}. Uranyl nitrate was irradiated by the Los Alamos Water Boiler Reactor and $^{147}$Pr and $^{148}$Pr were identified by $\beta$-counting and by measuring $\gamma$- and $\beta$-ray spectra following chemical separation. $^{149}$Pr was produced by irradiating enriched $^{150}$Nd with 22-24~MeV betatron bremsstrahlung. ``The following new $\beta$-decay chains of cerium and praseodymium were identified in the fission products of $^{235}$U: $^{147}$Ce(65$\pm$6 sec)-$^{147}$Pr(12.0$\pm$0.5 min) and $^{148}$Ce(43$\pm$10 sec)-$^{148}$Pr(1.98$\pm$0.10 min)... The mass assignments of 5.98 hr $^{145}$Pr and the 12 min $^{147}$Pr were substantiated by ($\gamma$,p) reactions on enriched $^{146}$Nd and $^{148}$Nd. Similar irradiations of $^{150}$Nd produced some evidence for an $\sim$2.3 min activity attributable to $^{149}$Pr.'' The half-lives of 12.0(5)~min for $^{147}$Pr, 1.98(10)~min for $^{148}$Pr, and 2.3~min for $^{149}$Pr agree with the presently adopted values of 13.4(4)~min, 2.29(2)~min (or the 2.01(7)~min isomeric state), and 2.26(7)~min, respectively. The 12~min half-life had previously been incorrectly assigned to $^{148}$Pr \cite{1960Wil02}.

\subsection*{$^{150}$Pr}\vspace{0.0cm}
``New isotope $^{150}$Pr'' was published in 1970 by Ward et al. documenting their observation of $^{150}$Pr \cite{1970War02}. A sample of enriched $^{150}$Nd was bombarded with 14.7~MeV neutrons produced by the $^3$H(d,n)$^4$He reaction from the Arkansas 400-kV Cockroft-Walton accelerator. ``In the present investigation, a new 6.1-sec $\beta^-$- and $\gamma$-ray activity was observed from the fast-neutron bombardment of enriched $^{150}$Nd metal and was assigned to $^{150}$Pr based on its decay characteristics.'' The observed half-life of 6.1(3)~s is in agreement with the currently accepted half-life of 6.19(16)~s. Previous observation of 66~keV and 74~keV $\gamma$ rays \cite{1970Wat01} were evidently incorrect.

\subsection*{$^{151}$Pr}\vspace{0.0cm}
Graefenstedt et al. published their observation of $^{151}$Pr in the 1990 paper titled ``Beta decay energies and nuclear masses of  $^{148}$Ba, $^{148}$La, and $^{151}$Pr'' \cite{1990Gra01}. $^{151}$Pr was produced in thermal fission of $^{239}$Pu at the Institute Laue-Langevin in Grenoble, France, and identified with the mass separator LOHENGRIN. ``For $^{151}$Pr, the $\beta$-spectra coincident with 9 different $\gamma$-transitions could be evaluated. From the endpoint energies given in [the table], a $\beta$-decay energy Q$_\beta$($^{151}$Pr)=4170$\pm$75~keV is obtained.'' The authors did not consider their observation a new discovery because of previous work published in a conference proceeding \cite{1978Pin01}. Less than three month later the half-life was reported independently \cite{1990And02}.

\subsection*{$^{152}$Pr}\vspace{0.0cm}
$^{152}$Pr was discovered in 1983 by Hill et al. and the results were published in the paper: ``Decay of neutron-rich $^{152}$Pr and $^{152}$Ce'' \cite{1983Hil01}. $^{152}$Pr was produced in thermal fission of $^{235}$U in the Brookhaven High Flux Beam Reactor and identified with the mass separator TRISTAN. ``The $^{152}$Pr half-life was measured to be 3.24$\pm$0.19~s, and its decay scheme was constructed based on $\gamma$ singles and coincidence measurements''. This half-life is close to the currently accepted value of 3.63(12)~s.

\subsection*{$^{153}$Pr}\vspace{0.0cm}
$^{153}$Pr was discovered in 1987 by Greenwood in the paper ``Identification of new neutron-rich rare-earth isotopes produced in $^{252}$Cf fission'' \cite{1987Gre01}. Spontaneous fission fragments from a $^{252}$Cf source were measured with the isotope separation on line (ISOL) system at the Idaho National Engineering Laboratory. $^{153}$Pr was identified by mass separation and the measurement of K x-rays. ``Identification of the $^{153}$Pr isotope was accomplished in two separate experiments, with collection-counting cycle times of 12 and 16 s each. The half-life value was obtained from an average of individual values involving the Nd K x rays and the 50.0-, 141.8-, and 191.8-KeV $\gamma$ rays. Separate half-life values from the x rays and the $\gamma$ rays are in excellent agreement, being 4.4 and 4.2 s, respectively.'' The observed half-life of 4.3(2)~s agrees with the currently accepted value of 4.28(11)~s. The observation of a 191.7~keV $\gamma$ ray in spontaneous $^{252}$Cf fission \cite{1972Hop01} were evidently incorrect \cite{2010Hwa01}.

\subsection*{$^{154}$Pr}\vspace{0.0cm}
``Identification of a new isotope $^{154}$Pr''  by Kawase and Okano described the first observation of $^{154}$Pr in 1988 \cite{1988Kaw01}. A $^{235}$U target was irradiated with neutrons from the Kyoto reactor and $^{154}$Pr was identified by measuring $\gamma$ rays following on-line mass separation. ``The Nd-K X-rays and 10 $\gamma$-rays have been assigned to be generated by the $\beta$-decay of $^{154}$Pr. The half-life had been determined to be 2.3(1)~s, which is not far from the theoretical predictions of 1.5~s and 4.0~s.''. This half-live corresponds to the currently accepted value.

\section{Discovery of $^{128-158}$Pm}

The element promethium was discovered in 1947 by Marinsky et al. \cite{1947Mar01}. In 1948 Marinsky and Glendenin suggested the name prometheum \cite{1948Mar01}. In this article previous claims for example suggesting the names florentium and illinium are discussed. In 1949, at the 15$^{th}$ Conference of the International Union of Chemistry \cite{1949IUC01} the name promethium was adopted to make the ending conform with that of other metals.

Thirty-one promethium isotopes from A = 128--258 have been discovered so far; these include 19 neutron-deficient and 12 neutron-rich isotopes. According to the HFB-14 model \cite{2007Gor01,2007HFB01}, $^{186}$Pm should be the last odd-odd particle stable neutron-rich nucleus while the odd-even particle stable neutron-rich nuclei should continue through $^{199}$Pm. The proton dripline has most likely been reached at $^{128}$Pm, however, eight more isotopes $^{120-127}$Pm could possibly still have half-lives longer than 10$^{-9}$~ns \cite{2004Tho01}. Thus, about 43 isotopes have yet to be discovered corresponding to 58\% of all possible promethium isotopes.

\begin{figure}
	\centering
	\includegraphics[scale=.7]{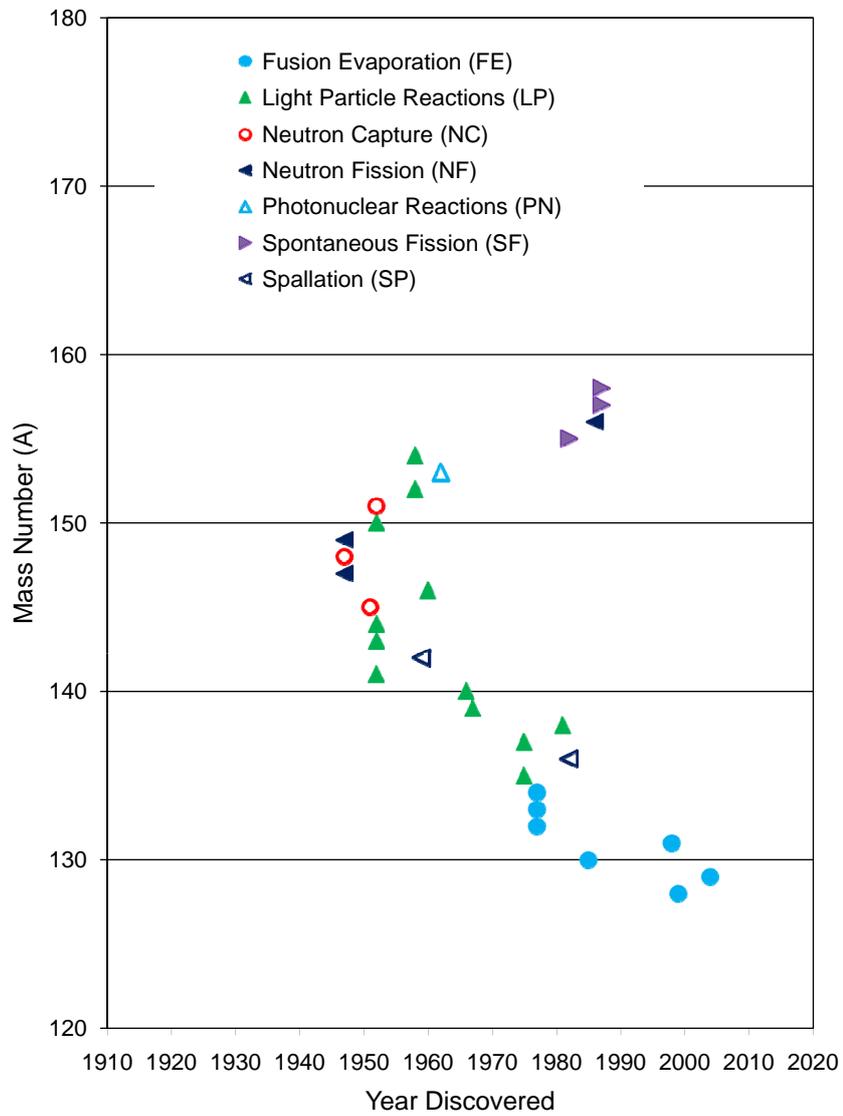}
	\caption{Promethium isotopes as a function of time when they were discovered. The different production methods are indicated.}
\label{f:year-pm}
\end{figure}

Figure \ref{f:year-pm} summarizes the year of first discovery for all promethium isotopes identified by the method of discovery. The range of isotopes predicted to exist is indicated on the right side of the figure. The radioactive promethium isotopes were produced using fusion evaporation reactions (FE), light-particle reactions (LP), neutron capture (NC), photo-nuclear (PN), neutron induced fission (NF), spontaneous fission (SF), and spallation reactions (SP). In the following, the discovery of each promethium isotope is discussed in detail.

\subsection*{$^{128}$Pm}\vspace{0.0cm}
Xu et al. first identified $^{128}$Pm in 1999 and reported the results in ``New $\beta$-delayed proton precursors in the rare-earth region near the proton drip line'' \cite{1999Xu01}. A 174~MeV $^{36}$Ar beam was accelerated with the sector-focused cyclotron at the National Laboratory of Heavy-Ion Accelerator in Lanzhou, China, and bombarded an enriched $^{96}$Ru target. Proton-$\gamma$ coincidences were measured in combination with a He-jet type transport system. ``From the decay curve the half-life of $^{128}$Pm was extracted to be 1.0$\pm$0.3~s'' This half-life is the currently accepted half-life value.

\subsection*{$^{129}$Pm}\vspace{0.0cm}
$^{129}$Pm was observed in 2004 by Xu et al. and published in the paper ``First decay study of the new isotope $^{129}$Pm near the proton drip line'' \cite{2004Xu01}. A 232~MeV $^{40}$Ca beam from the sector-focusing cyclotron in the Institute of Modern Physics in Lanzhou, China bombarded an enriched $^{92}$Mo target to form $^{129}$Pm in the fusion-evaporation reaction $^{92}$Mo($^{40}$Ca,p2n). X-rays and $\gamma$-rays were measured at the end of a helium jet transport system. ``The decay curve of the 99 keV $\gamma$ line gated on Nd-K$_\alpha$-X is shown in [the figure], from which the half-life of the $^{129}$Pm decay was extracted to be 2.4(9)~s.'' This half-life is the currently accepted half-life value.

\subsection*{$^{130}$Pm}\vspace{0.0cm}
$^{130}$Pm was discovered in 1985 by Wilmarth et al. and the results were published in the paper entitled ``Beta-delayed proton precursors with 59 $\leqslant$ Z $\leqslant$ 62'' \cite{1985Wil01}. A 170~MeV $^{40}$Ca beam from the Berkeley Super HILAC was used to produce $^{130}$Pm in the fusion-evaporation reaction $^{92}$Mo($^{40}$Ca,pn). Charged-particles, X-rays and $\gamma$-rays were measured following mass separation with the on-line separator OASIS. ``A weak $\beta$-delayed proton activity with a half life of 2.2$^{+0.6}_{-0.4}$s was observed at this mass. The Nd K x-rays measured in coincidence with the protons assign the activity to the new isotope $^{130}$Pm. This is the first observation of proton emission from a promethium precursor.'' The quoted half-life agrees with the currently accepted value of 2.6(2)~s.

\subsection*{$^{131}$Pm}\vspace{0.0cm}
``Yrast structures in the neutron-deficient $^{127}_{59}$Pr$_{69}$ and $^{131}_{61}$Pm$_{70}$ nuclei'' was published in 1998 by Parry et al. documenting their observation of $^{131}$Pm \cite{1998Par01}. $^{131}$Pm was produced in the fusion evaporation reactions $^{96}$Ru($^{40}$Ca,$\alpha$p) and $^{96}$Ru($^{39}$K,2p2n) at beam energies of 195~MeV and 200~MeV, respectively, at the TASCC facility at Chalk River Laboratory, Canada. Gamma-rays and charged particles were measured with the 8$\pi$ detector array and the ALF miniball. ``From systematics we assign the new bands as the $\pi$h$_{11/2}$ yrast bands of $^{131}$Pm and $^{127}$Pr, respectively.''

\subsection*{$^{132-134}$Pm}\vspace{0.0cm}
The first identification of $^{132}$Pm, $^{133}$Pm, and $^{134}$Pm was reported in 1977 by Bogdanov et al. in ``New neutron-deficient isotopes of barium and rare-earth elements'' \cite{1977Bog01}. The Dubna U-300 Heavy Ion Cyclotron accelerated a $^{32}$S beam which bombarded enriched targets of $^{102}$Pd and $^{106}$Cd. The isotopes were identified with the BEMS-2 isotope separator. ``In the present paper, isotopes were mainly identified by measuring the $\gamma$-ray and X-ray spectra of the daughter nuclei formed as a result of measuring the $\beta^+$ decay.'' The reported half-lives of 4(2)~s for $^{132}$Pm, 12(3)~s for $^{133}$Pm, and 24(2)~s for $^{134}$Pm are close to the accepted values of 6.3(7)~s, 15(3)~s, and 22(1)~s, respectively.

\subsection*{$^{135}$Pm}\vspace{0.0cm}
$^{135}$Pm was first observed by van Klinken and Feenstra in 1975 as reported in ``Shape implications of unhindered 11/2$^-$ $\rightarrow$ 11/2$^-$ $\beta$ decays in the region with N$<$82 and Z$>$50'' \cite{1975Van01}. Alpha beams accelerated to 140 MeV by the KVI cyclotron bombarded natural praseodymium targets. Gamma- and beta-rays were measured at the end of a pneumatic transport system. ``The isotope $^{135}$Pm, not reported before, was identified by its $\beta$ decay to the known level at 198.7~keV in $^{135}$Nd. It has a half-life of 44$\pm$9~s and was produced by $^{141}$Pr(140~MeV $\alpha$,10n)$^{135}$Pm.'' This half-life is consistent with the currently accepted value of 49(3)~s ground state or the 40(3)~s isomeric state.

\subsection*{$^{136}$Pm}\vspace{0.0cm}
The 1982 paper ``New neutron deficient isotopes with mass numbers A=136 and 145'' by Alkhazov et al. described the observation of $^{136}$Pm \cite{1982Alk01}. Tungsten and tantalum targets were bombarded with 1~GeV protons and $^{136}$Pr was produced in spallation reactions. Charged particles, X-rays, and $\gamma$-rays were measured at the end of a magnetic mass separator. ``The decays of a new 114.5~keV gamma-line (T$_{1/2}$=42$\pm$4~s) and 373.5~keV in a daughter isotope $^{136}$Pm are also seen in this figure.'' This half-life agrees with the currently accepted value of 47(2)~s. Alkhazov et al. did not consider their work as a new discovery referring to the Nuclear Data compilation \cite{1979Pek01}, which, however, was based on a private communication.

\subsection*{$^{137}$Pm}\vspace{0.0cm}
Nowicki et al. reported the observation of $^{137}$Pm in the 1975 paper ``The level scheme of $^{137}$Nd from ($\alpha$, xn) reactions and from $\beta$-decay of the 11/2$^{-}$ isomer of $^{137}$Pm'' \cite{1975Now01}. A 104 MeV $\alpha$-particle beam from the Karlsruhe isochronous cyclotron was used to irradiate a praseodymium target and $^{137}$Pm was formed in the reaction $^{141}$Pr($\alpha$,8n). $\gamma$-ray singles and coincidences were measured after an electromagnetic mass separator. ``After irradiation of $^{141}$Pr with 104 MeV $\alpha$-particles the $\gamma$-rays following the $^{137m}$Nd decay have been observed with a half life of 2.4$\pm$0.1 min and with intensity ratios different from those of the isomeric decay. As a 2.4~min activity is not found after irradiating $^{140}$Ce with $\alpha$-particles, this activity was associated to the decay of a $^{137}$Pm level.'' This half-life is the currently accepted value.

\subsection*{$^{138}$Pm}\vspace{0.0cm}
``Structure of even-even $^{138}$Nd from the decay $^{138}$Pm'' by Deslauriers et al. reported the observation of $^{138}$Pm in 1981 \cite{1981Des01}.  Nd$_{2}$O$_{3}$ targets enriched in $^{142}$Nd were bombarded with 35$-$80~MeV protons and $^{138}$Pm was formed in the reactions $^{142}$Nd(p,5n). Conversion electron spectra, as well as $\gamma$- and X-rays were measured. ``The decay half-life of $^{138}$Pm was measured accurately by the following time decay of the intense 437.4, 493.1, 520.9, 729.0, 1,279.1 keV gamma rays over a period of about 5 half-lives. The data were corrected for electronic dead time effects using the constant rate pulser technique and an average value of 3.24$\pm$0.05 m was obtained for the half-life of $^{138}$Pm. This is in agreement with the value of 3.5$\pm$0.3 m measured by van Klinken et al. \cite{1973Van01}.''. This half-life corresponds to the presently adopted value of 3.24(5)~min for the isomeric state. The van Klinken reference mentioned in the quote is only an internal report and was not published in a refereed journal. The ground state was first observed two years later \cite{1983Alk01}.

\subsection*{$^{139}$Pm}\vspace{0.0cm}
In 1967 Bleyl et al. reported the observation of $^{139}$Pm in ``\"Uber den Zerfall von $^{139}$Pm, $^{140}$Pm und $^{141}$Pm'' \cite{1967Ble01}. Enriched $^{142}$Nd was bombarded with 50~MeV deuterons from the Karlsruhe isochronous cyclotron. Beta-decay curves and $\gamma$-ray spectra were measured following chemical separation. ``The numerical analyses of the decay curve of the chemically separated Pm-sample yielded the following half-lives: $^{139}$Pm approximately 6~min...'' This half-life is consistent with the presently adopted value of 4.15(5)~min.

\subsection*{$^{140}$Pm}\vspace{0.0cm}
The discovery of $^{140}$Pm was published in the 1966 paper ``Promethium-140'' by Aten and Kapteijn \cite{1966Ate01}. $^{140}$Pm was produced in irradiations of $^{142}$Nd with 50 and 40 MeV protons and $^{144}$Sm with 50~MeV protons. Gamma-ray spectra were measured following chemical separation. ``We have already observed the formation of this activity by the irradiation of separated $^{142}$Nd with 50 and 40 MeV protons.. After irradiation with 17 MeV protons the activity was not observed. The data suggest that in this case we are dealing with $^{140}$Pm, formed by the reactions $^{142}$Nd(p,3n)$^{140}$Pm and $^{144}$Sm(p,$\alpha$n)$^{140}$Pm.'' The reported half-life of 6(1)~min agrees with the currently accepted value of 5.95(5)~min.

\subsection*{$^{141}$Pm}\vspace{0.0cm}
``Promethium isotopes'' was published in 1952 by Kistiakowsky documenting the observation of $^{141}$Pm \cite{1952Kis01}. Neodymium oxide enriched in $^{142}$Nd was bombarded with 20~MeV and 32~MeV protons from the Berkeley linear accelerator. Gamma- and beta-ray spectra were measured following chemical separation. ``The (Nd$^{142}$)$_{2}$O$_{3}$ yielded a 20-minutes half-life activity when bombarded with 20- and 32-MeV protons. This was assigned to Pm$^{141}$ from the Nd$^{142}$(p,2n) reaction.'' The measured half-life of 20(2)~min is in agreement with the currently accepted half life of 20.90(5)~min.

\subsection*{$^{142}$Pm}\vspace{0.0cm}
In 1959 Gratot et al. reported the observation of $^{142}$Pm in ``\'Etude de quelques isotopes tr\`es d\'eficients en neutrons du prom\'eth\'eum et du samarium'' \cite{1959Gra01}. Protons were accelerated with the Saclay cyclotron to 11 MeV and bombarded an enriched $^{142}$Nd target. Beta-decay curves were measured with a Geiger counter. ``Pour obtenir l'activit\'e $\beta^+$ de $^{142}$Pm, nous utilisons le fait que ces $\beta^+$ ont une \'energie de 3.78 MeV, alors que ceux de $^{17}$F ont une \'energie de 1.7 MeV.'' [In order to obtain the $\beta^+$ activity of $^{142}$Pm, we utilized the fact that these $\beta^+$ had an energy of 3.78~MeV, while those of $^{17}$F had an energy of 1.7~MeV.] The measured half-life of 40(5)~s agrees with the presently adopted value of 40.5(5)~s. Previously, Kistiakiowsky was only able to establish limits of $<$2~min or $>$200~y \cite{1952Kis01}. The 1958 Table of Isotope \cite{1958Str01} had assigned a 30~s half-life to $^{142}$Pm based on a private communication.

\subsection*{$^{143,144}$Pm}\vspace{0.0cm}
``Promethium isotopes'' was published in 1952 by Kistiakowsky documenting the observation of $^{143}$Pm and $^{144}$Pm \cite{1952Kis01}. Neodymium oxide enriched in $^{143}$Nd and $^{144}$Nd were bombarded with 8.9~MeV protons from the Berkeley 60-inch cyclotron. Gamma- and beta-ray spectra were measured following chemical separation. ``Bombardments of (Nd$^{143}$)$_{2}$O$_{3}$ and (Nd$^{144}$)$_{2}$O$_{3}$ separately with 8.9~Mev protons both yielded activities with half-lives of 200$-$400 days, consisting principally of electromagnetic radiation. These activities from the (Pr$^{141}$)$_6$O$_{11}$, (Nd$^{143}$)$_{2}$O$_{3}$, and (Nd$^{144}$)$_{2}$O$_{3}$ bombardments were examined on a scintillation spectrometer. The conclusions drawn from these measurements were that Pm$^{143}$ and Pm$^{144}$ are probably very similar in their decay characteristics and that the activity observed from (Pr$^{141}$)$_{6}$O$_{11}$ bombardments is probably a mixture due to both of these nuclides.'' These half-lives are consistent with the currently adopted values of 265(7)~d and 363(14)~d for $^{143}$Pm and $^{144}$Pm, respectively.

\subsection*{$^{145}$Pm}\vspace{0.0cm}
$^{145}$Pm was first observed in 1951 by Butement as described in ``Radioactive samarium-145 and promethium-145'' \cite{1951But03}. Neutron irradiation of a very pure samarium oxide sample was used to produce $^{145}$Sm which then decayed to $^{145}$Pm. Characteristic K X-rays were measured following chemical separation. ``It is concluded that samarium-145 decays by orbital electron capture, with a half-life of 410 days, into promethium-l45, which also decays by orbital electron capture into stable neodymium-145. By measuring the relative K X-ray activities of the
samarium-145, and the promethium-145 descended from it in a known time, and assuming equal counting efficiencies for these two X-rays of nearly equal
energies, the half-life of promethium-145 was calculated to be approximately thirty years.'' This half-life is within a factor of two of the presently adopted value 17.7(4)~y.

\subsection*{$^{146}$Pm}\vspace{0.0cm}
Funk et al. observed $^{146}$Pm in 1960 as described in ``Radioactive decay of Pm$^{143}$, Pm$^{144}$, and Pm$^{146}$'' \cite{1960Fun01}. An enriched $^{148}$Nd target was irradiated with protons from the Oak Ridge 86-in. cyclotron. $^{146}$Pm was produced by (p,3n) reactions and identified by measuring $\gamma$-ray and internal conversion electron spectra. ``$\gamma-\gamma$ coincidence measurements established that the 453-kev photons coincide with photons of about 745 kev. Coincidences obtained by gating with K x rays and sweeping the gamma-ray spectrum indicated that the 453-kev photons and some of the 745-kev photons coincide strongly with K x rays as shown in [the figure]. The number of K x ray-745-kev gamma ray coincidences was far too large to be accounted for by the Pm$^{143}$ present in the source. On the basis of these coincidence data, it was concluded that levels of 453 and 1198 kev in Nd$^{146}$ are populated by the electron-capture decay of Pm$^{146}$.'' The measured half-life of 710(70)~d is significantly shorter than the currently accepted value of 5.53(5)~y, however, the assigned $\gamma$-transitions of the daughter nucleus $^{146}$Nd are correct. A previously reported $\sim$1~y half-life of $^{146}$Pm \cite{1952Kis01} was not credited with the discovery because of the incorrect half-life lacking any additional supporting data.

\subsection*{$^{147}$Pm}\vspace{0.0cm}
In ``The chemical identification of radioisotopes of neodymium and of element 61'', Marinsky et al. reported the discovery of $^{147}$Pm in 1947 \cite{1947Mar01}. Fission fragments from $^{235}$U were analyzed as part of the Manhattan Project. $^{147}$Pm was identified by measuring X-, $\gamma$- and $\beta$-ray spectra following chemical separation. ``Radiochemical studies and fractionations have established the decay characteristics, genetic relation, and mass assignments of 11~d Nd$^{147}$, 3.7~y 61$^{147}$, 1.7~h Nd$^{(149)}$, and 47~h 61$^{149}$.'' The 3.7~h half-life for  $^{147}$Pm is close to the currently accepted value of 2.6234(2)~h. Although the ``verification'' of the mass assignment by Inghram et al. \cite{1947Ing04} was submitted about three months earlier, Marinsky et al. is credited with the discovery because as part of the Manhattan Project Inghram et al. were aware of the earlier work by Marinsky et al.

\subsection*{$^{148}$Pm}\vspace{0.0cm}
$^{148}$Pm was first observed in 1947 by Parker et al. as reported in ``The 5.3 day isotope in element 61'' \cite{1947Par02}. A promethium sample consisting predominantly of $^{147}$Pm was irradiated by slow neutrons in the Clinton Pile. ``Upon development this large plate showed equal blackening at mass 148 and increasing blackening for successive transfers at mass 147. Thus the active isotope in element 61 with a half-life of 5.3 days is at mass 148.'' This half-life is in agreement with the currently accepted value of 5.368(2)~d.

\subsection*{$^{149}$Pm}\vspace{0.0cm}
In ``The chemical identification of radioisotopes of neodymium and of element 61'', Marinsky et al. reported the discovery of $^{149}$Pm in 1947 \cite{1947Mar01}. Fission fragments from $^{235}$U were analyzed as part of the Manhattan Project. $^{149}$Pm was identified by measuring X-, $\gamma$- and $\beta$-ray spectra following chemical separation. ``Radiochemical studies and fractionations have established the decay characteristics, genetic relation, and mass assignments of 11~d Nd$^{147}$, 3.7~y 61$^{147}$, 1.7~h Nd$^{(149)}$, and 47~h 61$^{149}$.'' The 47~h half-life for  $^{149}$Pm is in agreement with the currently accepted value of 53.08(5)~h. Although the ``verification'' of the mass assignment by Inghram et al. \cite{1947Ing04} was submitted about three months earlier, Marinsky et al. is credited with the discovery because as part of the Manhattan Project Inghram et al. were aware of the earlier work by Marinsky et al.

\subsection*{$^{150}$Pm}\vspace{0.0cm}
``Radioactive Pm$^{148}$ and Pm$^{150}$'' was published in 1952 by Long and Pool reporting their observation of $^{150}$Pm \cite{1952Lon01}. Enriched neodymium oxide targets were bombarded with 6-MeV protons and decay curves were measured with a Geiger counter and Wulf electrometer. ``The 2.7-hour activity may therefore, be assigned to Pm$^{150}$ in accordance with the reaction Nd$^{150}$(p,n)Pm$^{150}$'' This half-life agrees with the presently accepted value of 2.68(2)~h.

\subsection*{$^{151}$Pm}\vspace{0.0cm}
$^{151}$Pm was identified in 1952 by Rutledge et al. and published in ``Gamma-rays associated with selected neutron-induced radioactivities'' \cite{1952Rut01}. Enriched Nd$^{150}$ was bombarded with neutrons in the Argonne heavy water moderated reactor. $^{151}$Pm was identified with help of a photographic magnetic spectrometer. ``A newly discovered daughter product of Nd$^{151}$ is found to have a half-life of 27.5$\pm$1.5 hours. The assignment of this activity to Pm$^{151}$ is based upon the facts that it is produced by bombarding enriched Nd$^{150}$ with neutrons, and the internal conversion lines associated with the fifteen observed gamma-rays show work function differences of samarium.'' The quoted half-life is in agreement with the currently accepted value of 28.40(4)~h. A tentative assignment of a 12~min activity to $^{151}$Pm \cite{1951Mar01} were evidently incorrect.

\subsection*{$^{152}$Pm}\vspace{0.0cm}
$^{152}$Pm was discovered by Wille and Fink in the 1958 paper ``Two new promethium isotopes; cross sections of some samarium isotopes for 14.8-MeV neutrons'' \cite{1958Wil01}. Enriched $^{152}$Sm$_{2}$O$_3$ was bombarded with 14.8-MeV neutrons produced in the reaction $^3$H(d,n)$^4$He from the Arkansas Cockroft-Walton accelerator. Decay and absorption curves were measured. ``When highly enriched samples of Sm$^{152}$ and Sm$^{154}$ are irradiated with 14.8-MeV neutrons, activities having half-lives of 6.5$\pm$0.5 min and 2.5 $\pm$0.5 min are observed. On the basis of yields and cross bombardments, these are assigned to new isotopes Pm$^{152}$ and Pm$^{154}$, respectively.'' The quoted half-life for $^{152}$Pm of 6.5(5)~min is close to the currently accepted value of 7.52(8)~min for an isomeric state. The ground state was first observed eleven years later \cite{1969Wak01}.

\subsection*{$^{153}$Pm}\vspace{0.0cm}
The 1962 paper ``Decay of a new nuclide promethium 153'' by Kotajima described the first observation of $^{153}$Pm \cite{1962Kot01}. A sample of $^{154}$Sm$_{2}$O$_{3}$ was irradiated with 20~MeV bremsstrahlung at the JAERI linear accelerator. Beta- and gamma-rays singles as well as coincidences were recorded. ``A new nuclide promethium 153 was investigated from the $^{154}$Sm($\gamma$,p)$^{153}$Pm reaction induced by 20 MeV bremsstrahlung. The activity was measured with beta- and gamma-ray scintillation spectrometers.  The half life of this nuclide was found to be T$_{1/2}$=5.5$\pm$0.2~min and the end point energy of the beta ray was found to be E$_{\beta}$ = 16.5$\pm$0.05~MeV.'' The quoted half-life is in agreement with the currently accepted half-life of 5.25(2)~min.

\subsection*{$^{154}$Pm}\vspace{0.0cm}
$^{154}$Pm was discovered by Wille and Fink in the 1958 paper ``Two new promethium isotopes; cross sections of some samarium isotopes for 14.8-MeV neutrons'' \cite{1958Wil01}. Enriched $^{154}$Sm$_{2}$O$_3$  was bombarded with 14.8-MeV neutrons produced in the reaction $^3$H(d,n)$^4$He from the Arkansas Cockroft-Walton accelerator. Decay and absorption curves were measured. ``When highly enriched samples of Sm$^{152}$ and Sm$^{154}$ are irradiated with 14.8-MeV neutrons, activities having half-lives of 6.5$\pm$0.5 min and 2.5 $\pm$0.5 min are observed. On the basis of yields and cross bombardments, these are assigned to new isotopes Pm$^{152}$ and Pm$^{154}$, respectively.'' The quoted half-life for $^{154}$Pm of 2.5(5)~min agrees with the currently accepted value of 2.68(7)~min.

\subsection*{$^{155}$Pm}\vspace{0.0cm}
``Identification of a new isotope, $^{155}$Pm, produced in $^{252}$Cf fission'' was published in 1982 by Greenwood et al. reporting their observation of $^{155}$Pm \cite{1982Gre01}. $^{155}$Pm was identified in spontaneous fission of $^{252}$Cf by high-performance liquid chromatography. ``The assignment of the 48-s activity to $^{155}$Pm is based in part upon the observations of the growth of 22.4-min $^{155}$Sm from a 48-s parent and in parton the fact that the 5 $\gamma$-rays could be placed in a $^{155}$Sm level scheme which is consistent with that obtained recently from studies of the $^{154}$Sm(n,$\gamma$) reaction.'' This half life is in agreement with the currently accepted value of 41.5(2)~s.

\subsection*{$^{156}$Pm}\vspace{0.0cm}
$^{156}$Pm was discovered in 1986 by Mach et al. with the results published in their paper titled ``Identification of four new neutron rare-earth isotopes'' \cite{1986Mac01}. $^{156}$Pm was produced in thermal neutron fission of $^{235}$U at Brookhaven National Laboratory. X-rays and $\gamma$-rays were measured at the on-line mass separator TRISTAN. ``Eleven transitions were found associated with the decay of $^{156}$Pm on the basis of strong x-$\gamma$, and $\gamma$-$\gamma$ coincidences. The half-life measurement represents the average of six results involving the 75.7-,
117.8-, and 174.1-keV transitions and two sets of GMS measurements with GMS cycle times of $\Delta$T = 48 and 96 sec.'' The measured half-life of 28.2(11)~s is in agreement with the currently accepted value of 26.70(10)~s. Less than a month later Okano et al. independently reported a half-life of 29(2)~s \cite{1986Oka02}.

\subsection*{$^{157,158}$Pm}\vspace{0.0cm}
In 1987, Greenwood et al. identified $^{157}$Pm and $^{158}$Pm in the paper entitled ``Identification of new neutron-rich rare-earth isotopes produced in $^{252}$Cf Fission'' \cite{1987Gre01}. Spontaneous fission fragments from a $^{252}$Cf source were measured with the isotope separation on line (ISOL) system at the Idaho National Engineering Laboratory. $^{157}$Pm and $^{158}$Pm were identified by mass separation and the measurement of K x-rays. ``$^{157}$Pm. Some 17 $\gamma$ rays could be associated with the decay of $^{157}$Pm in the present work. The half-life value was obtained as an average of individual values involving the Sm K x rays and the 52.6-, 108.2-, 160.5-, and 187.9-keV $\gamma$ rays. Separate half-life values for the x rays and $\gamma$ rays were identical. $^{158}$Pm. Identification of the $^{158}$Pm decay was accomplished in two separate experiments, with collection country cycle times of 12 and 16 s, respectively. One $\gamma$ ray, at 72.7 keV, together with the Sm K x rays could be associated with this activity. The half-life values determined for the x rays and 72.7-keV $\gamma$ rays were in reasonable agreement, being (4.5$\pm$0.5) s and (5.5$\pm$0.8) s, respectively.'' The observed half-life of 10.90(20)~s for $^{157}$Pm is in agreement with the currently accepted value of 10.56(10)s. The observed half-life of 4.8(5)~s corresponds to the presently adopted value for $^{158}$Pm.

\section{Summary}
The discoveries of the known cesium, lanthanum, praseodymium and promethium isotopes have been compiled and the methods of their production discussed. The identification of these isotopes was relatively straightforward with only a few initially incorrect half-lives reported. Most of the misidentifications occurred in the praseodymium isotopes.

No incorrect assignments were reported for the cesium isotopes and only the half-lives of five isotopes ($^{130}$Cs, $^{134}$Cs, and $^{138-140}$Cs) were initially reported without a mass assignment. The half-lives of the lanthanum isotopes 141 through 143 were reported before it was possible to make firm mass assignments and only the half-life of $^{136}$La was initially wrong. In contrast, for five praseodymium isotopes ($^{121}$Pr, $^{127}$Pr, $^{134}$Pr, $^{136}$Pr, and $^{137}$Pr) a wrong half-life was first reported and the half-lives of $^{142}$Pr and $^{144-146}$Pr were at first misidentified. For promethium only $^{151}$Pm was initially assigned an incorrect half-life, and the half-life of $^{147}$Pm was at first assigned to $^{148}$Pm.

\section*{Acknowledgements}

This work was supported by the National Science Foundation under grant No. PHY06-06007 (NSCL).

\bibliography{../isotope-discovery-references}

\newpage

\newpage

\TableExplanation

\bigskip
\renewcommand{\arraystretch}{1.0}

\section*{Table 1.\label{tbl1te} Discovery of cesium, lanthanum, praseodymium and promethium isotopes }
\begin{tabular*}{0.95\textwidth}{@{}@{\extracolsep{\fill}}lp{5.5in}@{}}
\multicolumn{2}{p{0.95\textwidth}}{ }\\

Isotope & Cesium, lanthanum, praseodymium and promethium isotope \\
Author & First author of refereed publication \\
Journal & Journal of publication \\
Ref. & Reference \\
Method & Production method used in the discovery: \\

  & FE: fusion evaporation \\
  & LP: light-particle reactions (including neutrons) \\
  & MS: mass spectroscopy \\
  & NC: neutron capture reactions \\
  & PN: photo-nuclear reactions \\
  & NF: neutron induced fission \\
  & CPF: charged-particle induced fission \\
  & SF: spontaneous fission \\
  & SP: spallation \\
  & PF: projectile fragmentation of fission \\

Laboratory & Laboratory where the experiment was performed\\
Country & Country of laboratory\\
Year & Year of discovery \\
\end{tabular*}
\label{tableI}

\datatables 



\setlength{\LTleft}{0pt}
\setlength{\LTright}{0pt}


\setlength{\tabcolsep}{0.5\tabcolsep}

\renewcommand{\arraystretch}{1.0}

\footnotesize 

\begin{longtable}{@{\extracolsep\fill}llllllll@{}}
\caption[Discovery of cesium, lanthanum, praseodymium and promethium isotopes.]{Discovery of cesium, lanthanum, praseodymium and promethium isotopes. See page\ \pageref{tbl1te} for Explanation of Tables}
Isotope & Author & Journal & Ref. & Method & Laboratory & Country & Year\\
\hline\\
\endfirsthead\\
\caption[]{(continued)}
Isotope & Author & Journal & Ref. & Method & Laboratory & Country & Year\\
\hline\\
\endhead
$^{112}$Cs & R.D. Page & Phys. Rev. Lett. &\cite{1994Pag01}& FE & Daresbury & UK &1994 \\
$^{113}$Cs & T. Faestermann & Phys. Lett. B &\cite{1984Fae01}& FE & Munich & Germany &1984 \\
$^{114}$Cs & J.M. D'Auria & Nucl. Phys. A &\cite{1978DAu01}& SP & CERN & Switzerland &1978 \\
$^{115}$Cs & J.M. D'Auria & Nucl. Phys. A &\cite{1978DAu01}& SP & CERN & Switzerland &1978 \\
$^{116}$Cs & D.D. Bogdanov & Sov. J. Nucl. Phys. &\cite{1975Bog01}& FE & Dubna & Russia &1975 \\
$^{117}$Cs & H. L. Ravn & Phys. Lett. B &\cite{1972Rav01}& SP & CERN & Switzerland &1972 \\
$^{118}$Cs & J. Chaumont & Phys. Lett. B &\cite{1969Cha01}& SP & CERN & Switzerland &1969 \\
$^{119}$Cs & J. Chaumont & Phys. Lett. B &\cite{1969Cha01}& SP & CERN & Switzerland &1969 \\
$^{120}$Cs & J. Chaumont & Phys. Lett. B &\cite{1969Cha01}& SP & CERN & Switzerland &1969 \\
$^{121}$Cs & J. Chaumont & Phys. Lett. B &\cite{1969Cha01}& SP & CERN & Switzerland &1969 \\
$^{122}$Cs & J. Chaumont & Phys. Lett. B &\cite{1969Cha01}& SP & CERN & Switzerland &1969 \\
$^{123}$Cs & H.B. Mathur & Phys. Rev. &\cite{1954Mat01}& LP & Berkeley & USA &1954 \\
$^{124}$Cs & J. Chaumont & Phys. Lett. B &\cite{1969Cha01}& SP & CERN & Switzerland &1969 \\
$^{125}$Cs & M.C. Michel & Phys. Rev. &\cite{1954Mic01}& LP & Berkeley & USA &1954 \\
$^{126}$Cs & M.I. Kalkstein & J. Inorg. Nucl. Chem. &\cite{1954Kal01}& FE & Berkeley & USA &1954 \\
$^{127}$Cs & R.W. Fink & Phys. Rev. &\cite{1950Fin02}& LP & Berkeley & USA &1950 \\
$^{128}$Cs & R.W. Fink & J. Inorg. Nucl. Chem. &\cite{1951Fin01}& LP & Berkeley & USA &1951 \\
$^{129}$Cs & R.W. Fink & Phys. Rev. &\cite{1950Fin02}& LP & Berkeley & USA &1950 \\
$^{130}$Cs & A.B. Smith & Phys. Rev. &\cite{1952Smi01}& LP & Indiana & USA &1952 \\
$^{131}$Cs & F. Yu & Phys. Rev. &\cite{1947Yu01}& NC & Ohio State & USA &1947 \\
$^{132}$Cs & A.H. Wapstra & Physica &\cite{1953Wap01}& LP & Amsterdam & Netherlands &1953 \\
$^{133}$Cs & F.W. Aston & Nature &\cite{1921Ast03}& MS & Cambridge & UK &1921 \\
$^{134}$Cs & D.C. Kalbfell & Phys. Rev. &\cite{1940Kal01}& NC & Berkeley & USA &1940 \\
$^{135}$Cs & N. Sugarman & Phys. Rev. &\cite{1949Sug02}& NF & Los Alamos & USA &1949 \\
$^{136}$Cs & L.E. Glendenin & Nat. Nucl. Ener. Ser. &\cite{1951Gle03}& NF & Oak Ridge & USA &1951 \\
$^{137}$Cs & A. Turkevich & Nat. Nucl. Ener. Ser. &\cite{1951Tur01}& LP & Argonne & USA &1951 \\
$^{138}$Cs & W. Seelmann-Eggebert & Naturwiss. &\cite{1943See03}& LP & Berlin & Germany &1943 \\
$^{139}$Cs & F. A. Heyn & Nature &\cite{1939Hey01}& NF & Eindhoven & Netherlands &1939 \\
$^{140}$Cs & N. Sugarman & J. Chem. Phys. &\cite{1950Sug01}& NF & Argonne & USA &1950 \\
$^{141}$Cs & A.C. Wahl & Phys. Rev. &\cite{1962Wah01}& NF & St. Louis & USA &1962 \\
$^{142}$Cs & K. Fritze & Can. J. Phys. &\cite{1962Fri01}& NF & McMaster & Canada &1962 \\
$^{143}$Cs & K. Fritze & Can. J. Phys. &\cite{1962Fri01}& NF & McMaster & Canada &1962 \\
$^{144}$Cs & I. Amarel & Phys. Lett. B &\cite{1967Ama01}& CPF & Orsay & France &1967 \\
$^{145}$Cs & B.L. Tracy & Phys. Lett. B &\cite{1971Tra01}& CPF & Grenoble & France &1971 \\
$^{146}$Cs & B.L. Tracy & Phys. Lett. B &\cite{1971Tra01}& CPF & Grenoble & France &1971 \\
$^{147}$Cs & F.K. Wohn & Phys. Rev. C &\cite{1978Woh01}& NF & Grenoble & France &1978 \\
$^{148}$Cs & E. Koglin & Z. Phys. A &\cite{1978Kog01}& NF & Grenoble & France &1978 \\
$^{149}$Cs & H.L. Ravn & Phys. Rep. &\cite{1979Rav01}& SP & CERN & Switzerland &1979 \\
$^{150}$Cs & H.L. Ravn & Phys. Rep. &\cite{1979Rav01}& SP & CERN & Switzerland &1979 \\
$^{151}$Cs & H.L. Ravn & Phys. Rep. &\cite{1979Rav01}& SP & CERN & Switzerland &1979 \\
$^{152}$Cs & H.L. Ravn & Phys. Rep. &\cite{1979Rav01}& SP & CERN & Switzerland &1979 \\
 & & & & & &  \\
 & & & & & &  \\
$^{117}$La & F. Soramel & Phys. Rev. C &\cite{2001Sor01}& FE & Legnaro & Italy &2001 \\
$^{118}$La &&&&&&& \\
$^{119}$La &&&&&&& \\
$^{120}$La & J.M. Nitschke & Nucl. Phys. A &\cite{1984Nit01}& FE & Berkeley & USA &1984 \\
$^{121}$La & T. Sekine & Z. Phys. A &\cite{1988Sek01}& FE & JAERI & Japan &1988 \\
$^{122}$La & J.M. Nitschke & Nucl. Phys. A &\cite{1984Nit01}& FE & Berkeley & USA &1984 \\
$^{123}$La & D.D. Bogdanov & Nucl. Phys. A &\cite{1978Bog01}& FE & Dubna & Russia &1978 \\
$^{124}$La & D.D. Bogdanov & Nucl. Phys. A &\cite{1978Bog01}& FE & Dubna & Russia &1978 \\
$^{125}$La & K.Nakai & Phys. Lett. B &\cite{1973Nak01}& FE & Berkeley & USA &1973 \\
$^{126}$La & R.K. Sheline & Phys. Rev. Lett. &\cite{1961She01}& FE & Berkeley & USA &1961 \\
$^{127}$La & I.L. Preiss & Phys. Rev. &\cite{1963Pre01}& FE & Yale & USA &1963 \\
$^{128}$La & R.K. Sheline & Phys. Rev. Lett. &\cite{1961She01}& FE & Berkeley & USA &1961 \\
$^{129}$La & I.L. Preiss & Phys. Rev. &\cite{1963Pre01}& FE & Yale & USA &1963 \\
$^{130}$La & R.K. Sheline & Phys. Rev. Lett. &\cite{1961She01}& FE & Berkeley & USA &1961 \\
$^{131}$La & M.M. Gransden & Phys. Rev. &\cite{1951Gra01}& LP & McGill & Canada &1951 \\
$^{132}$La & M.M. Gransden & Phys. Rev. &\cite{1951Gra01}& LP & McGill & Canada &1951 \\
$^{133}$La & R.A. Naumann & Phys. Rev. &\cite{1950Nau01}& LP & Berkeley & USA &1950 \\
$^{134}$La & B.J. Stover & Phys. Rev. &\cite{1951Sto01}& LP & Berkeley & USA &1951 \\
$^{135}$La & J.B. Chubbuck & Phys. Rev. &\cite{1948Chu01}& LP & Berkeley & USA &1948 \\
$^{136}$La & R.A. Naumann & Phys. Rev. &\cite{1950Nau01}& LP & Berkeley & USA &1950 \\
$^{137}$La & M.G. Inghram & Phys. Rev. &\cite{1948Ing01}& LP & Argonne & USA &1948 \\
$^{138}$La & M.G. Inghram & Phys. Rev. &\cite{1947Ing01}& MS & Argonne & USA &1947 \\
$^{139}$La & F.W. Aston & Nature &\cite{1924Ast04}& MS & Cambridge & UK &1924 \\
$^{140}$La & J. K. Marsh & Nature &\cite{1935Mar01}& NC & Oxford & UK &1935 \\
$^{141}$La & S. Katcoff & Nat. Nucl. Ener. Ser. &\cite{1951Kat01}& NF & Argonne & USA &1951 \\
$^{142}$La & A. Vanden Bosch& Physica &\cite{1953Van01}& NF & Amsterdam & Netherlands &1953 \\
$^{143}$La & H. Gest & Nat. Nucl. Ener. Ser. &\cite{1951Ges01}& NF & Oak Ridge & USA &1951 \\
$^{144}$La & I. Amarel & Phys. Lett. B &\cite{1967Ama01}& CPF & Orsay & France &1967 \\
$^{145}$La & P.O. Aronsson & J. Inorg. Nucl. Chem. &\cite{1974Aro01}& NF & Chalmers & Sweden &1974 \\
$^{146}$La & R.L. Watson & Nucl. Phys. A &\cite{1970Wat01}& SF & Berkeley & USA &1970 \\
$^{147}$La & G. Engler & Phys. Rev. C &\cite{1979Eng01}& NF & Soreq & Israel &1979 \\
$^{148}$La & H. Gabelmann & Z. Phys. A &\cite{1982Gab01}& NF & Grenoble & France &1982 \\
$^{149}$La & G. Engler & Phys. Rev. C &\cite{1979Eng01}& NF & Soreq & Israel &1979 \\
$^{150}$La & G. Rudstam & At. Data Nucl. Data Tables &\cite{1993Rud01}& NF & Studsvik & Sweden &1993 \\
$^{151}$La & M. Bernas & Phys. Lett. B &\cite{1994Ber01}& PF & Darmstadt & Germany &1994 \\
$^{152}$La & M. Bernas & Phys. Lett. B &\cite{1994Ber01}& PF & Darmstadt & Germany &1994 \\
$^{153}$La & M. Bernas & Phys. Lett. B &\cite{1994Ber01}& PF & Darmstadt & Germany &1994 \\
 & & & & & &  \\
 & & & & & &  \\
$^{121}$Pr & A.P. Robinson & Phys. Rev. Lett. &\cite{2005Rob01}& FE & Argonne & USA &2005 \\
$^{122}$Pr &&&&&&& \\
$^{123}$Pr &&&&&&& \\
$^{124}$Pr & P.A. Wilmarth & Z. Phys. A &\cite{1986Wil01}& FE & Berkeley & USA &1986 \\
$^{125}$Pr & A.N. Wilson & Phys. Rev. C &\cite{2002Wil01}& FE & Argonne & USA &2002 \\
$^{126}$Pr & J.M. Nitschke & Z. Phys. A &\cite{1983Nit01}& FE & Berkeley & USA &1983 \\
$^{127}$Pr & A. Gizon & Z. Phys. A &\cite{1995Giz01}& FE & Grenoble & France &1995 \\
$^{128}$Pr & P.A. Wilmarth & Z. Phys. A &\cite{1985Wil01}& FE & Berkeley & USA &1985 \\
$^{129}$Pr & D.D. Bogdanov & Nucl. Phys. A &\cite{1977Bog01}& FE & Dubna & Russia &1977 \\
$^{130}$Pr & D.D. Bogdanov & Nucl. Phys. A &\cite{1977Bog01}& FE & Dubna & Russia &1977 \\
$^{131}$Pr & D.D. Bogdanov & Nucl. Phys. A &\cite{1977Bog01}& FE & Dubna & Russia &1977 \\
$^{132}$Pr & A. Latuszynski & Nukleonika &\cite{1974Lat02}& SP & Dubna & Russia &1974 \\
$^{133}$Pr & A.A. Abdurazakov & Bull. Acad. Sci. USSR &\cite{1970Abd01}& SP & Dubna & Russia &1970 \\
$^{134}$Pr & J.E. Clarkson & Nucl. Phys. A &\cite{1967Cla01}& FE & Berkeley & USA &1967 \\
$^{135}$Pr & T.H. Handley & Phys. Rev. &\cite{1954Han03}& LP & Oak Ridge & USA &1954 \\
$^{136}$Pr & Zh. Zhelev & Bull. Acad. Sci. USSR &\cite{1968Zhe01}& SP & Dubna & Russia &1968 \\
$^{137}$Pr & C. Dahlstrom & Can. J. Phys. &\cite{1958Dah01}& LP & McGill & Canada &1958 \\
$^{138}$Pr & B.J. Stover & Phys. Rev. &\cite{1951Sto01}& LP & Berkeley & USA &1951 \\
$^{139}$Pr & B.J. Stover & Phys. Rev. &\cite{1951Sto01}& LP & Berkeley & USA &1951 \\
$^{140}$Pr & M.L. Pool & Phys. Rev. &\cite{1938Poo02}& LP & Michigan & USA &1938 \\
$^{141}$Pr & F.W. Aston & Nature &\cite{1924Ast04}& MS & Cambridge & UK &1924 \\
$^{142}$Pr & J. K. Marsh & Nature &\cite{1935Mar01}& NC & Oxford & UK &1935 \\
$^{143}$Pr & M.L. Pool & Phys. Rev. &\cite{1948Poo01}& LP & Ohio State & USA &1948 \\
$^{144}$Pr & W.H. Burgus & Nat. Nucl. Ener. Ser. &\cite{1951Bur01}& NF & Argonne & USA &1951 \\
$^{145}$Pr & S.S. Markowitz & Phys. Rev. &\cite{1954Mar01}& NF & Brookhaven & USA &1954 \\
$^{146}$Pr & A.A. Caretto & Phys. Rev. &\cite{1953Car01}& NF & Brookhaven & USA &1953 \\
$^{147}$Pr & D.C. Hoffman & J. Inorg. Nucl. Chem. &\cite{1964Hof01}& NF & Los Alamos & USA &1964 \\
$^{148}$Pr & D.C. Hoffman & J. Inorg. Nucl. Chem. &\cite{1964Hof01}& NF & Los Alamos & USA &1964 \\
$^{149}$Pr & D.C. Hoffman & J. Inorg. Nucl. Chem. &\cite{1964Hof01}& PN & Los Alamos & USA &1964 \\
$^{150}$Pr & T.E. Ward & Phys. Rev. C &\cite{1970War02}& LP & Arkansas & USA &1970 \\
$^{151}$Pr & M. Graefenstedt & Z. Phys. A &\cite{1990Gra01}& NF & Grenoble & France &1990 \\
$^{152}$Pr & J.C. Hill & Phys. Rev. C &\cite{1983Hil01}& NF & Brookhaven & USA &1983 \\
$^{153}$Pr & R.C. Greenwood & Phys. Rev. C &\cite{1987Gre01}& SF & Idaho Falls & USA &1987 \\
$^{154}$Pr & Y. Kawase & Z. Phys. A &\cite{1988Kaw01}& NF & Kyoto & Japan &1988 \\
 & & & & & &  \\
 & & & & & &  \\
$^{128}$Pm& S.-W. Xu & Phys. Rev. C &\cite{1999Xu01}& FE & Lanzhou & China &1999 \\
$^{129}$Pm & S.-W. Xu & Eur. Phys. J. A &\cite{2004Xu01}& FE & Lanzhou & China &2004 \\
$^{130}$Pm & P.A. Wilmarth & Z. Phys. A &\cite{1985Wil01}& FE & Berkeley & USA &1985 \\
$^{131}$Pm & C.M. Parry & Phys. Rev. C &\cite{1998Par01}& FE & Chalk River & Canada &1998 \\
$^{132}$Pm & D.D. Bogdanov & Nucl. Phys. A &\cite{1977Bog01}& FE & Dubna & Russia &1977 \\
$^{133}$Pm & D.D. Bogdanov & Nucl. Phys. A &\cite{1977Bog01}& FE & Dubna & Russia &1977 \\
$^{134}$Pm & D.D. Bogdanov & Nucl. Phys. A &\cite{1977Bog01}& FE & Dubna & Russia &1977 \\
$^{135}$Pm & J. van Klinken & Phys. Rev. C &\cite{1975Van01}& LP & Groningen & Netherlands &1975 \\
$^{136}$Pm & G.D. Alkhazov & Z. Phys. A &\cite{1982Alk01}& SP & Dubna & Russia &1982 \\
$^{137}$Pm & G.P. Nowicki & Nucl. Phys. A &\cite{1975Now01}& LP & Karlsruhe & Germany &1975 \\
$^{138}$Pm & J. Deslauriers & Z. Phys. A &\cite{1981Des01}& LP & McGill & Canada &1981 \\
$^{139}$Pm & H.-J. Bleyl & Radiochim. Acta &\cite{1967Ble01}& LP & Karlsruhe & Germany &1967 \\
$^{140}$Pm & A.H.W. Aten & Physica &\cite{1966Ate01}& LP & Amsterdam & Netherlands &1966 \\
$^{141}$Pm & V. Kistiakowsky & Phys. Rev. &\cite{1952Kis01}& LP & Berkeley & USA &1952 \\
$^{142}$Pm & I. Gratot & Nucl. Phys. &\cite{1959Gra01}& SP & Orsay & France &1959 \\
$^{143}$Pm & V. Kistiakowsky & Phys. Rev. &\cite{1952Kis01}& LP & Berkeley & USA &1952 \\
$^{144}$Pm & V. Kistiakowsky & Phys. Rev. &\cite{1952Kis01}& LP & Berkeley & USA &1952 \\
$^{145}$Pm & F.D.S. Butement & Nature &\cite{1951But03}& NC & Harwell & UK &1951 \\
$^{146}$Pm & E.G. Funk Jr. & Phys. Rev. &\cite{1960Fun01}& LP & Oak Ridge & USA &1960 \\
$^{147}$Pm & J.A. Marinsky & J. Am. Chem. Soc. &\cite{1947Mar01}& NF & Oak Ridge & USA &1947 \\
$^{148}$Pm & G.W. Parker & Phys. Rev. &\cite{1947Par02}& NC & Oak Ridge & USA &1947 \\
$^{149}$Pm & J.A. Marinsky & J. Am. Chem. Soc. &\cite{1947Mar01}& NF & Oak Ridge & USA &1947 \\
$^{150}$Pm & J. K. Long & Phys. Rev. &\cite{1952Lon01}& LP & Ohio State & USA &1952 \\
$^{151}$Pm & W.C. Rutledge & Phys. Rev. &\cite{1952Rut01}& NC & Argonne & USA &1952 \\
$^{152}$Pm & R.G. Wille & Phys. Rev. &\cite{1958Wil01}& LP & Arkansas & USA &1958 \\
$^{153}$Pm & K. Kotajima & Nucl. Phys. &\cite{1962Kot01}& PN & JAERI & Japan &1962 \\
$^{154}$Pm & R.G. Wille & Phys. Rev. &\cite{1958Wil01}& LP & Arkansas & USA &1958 \\
$^{155}$Pm & R.C. Greenwood & Radiochim. Acta &\cite{1982Gre01}& SF & Idaho Falls & USA &1982 \\
$^{156}$Pm & H. Mach & Phys. Rev. Lett. &\cite{1986Mac01}& NF & Brookhaven & USA &1986 \\
$^{157}$Pm & R.C. Greenwood & Phys. Rev. C &\cite{1987Gre01}& SF & Idaho Falls & USA &1987 \\
$^{158}$Pm & R.C. Greenwood & Phys. Rev. C &\cite{1987Gre01}& SF & Idaho Falls & USA &1987 \\
 \\
\end{longtable}

\end{document}